\documentclass[structabstract]{aa}

\usepackage{graphicx}
\usepackage{epsfig}
\usepackage{natbib}
\usepackage[varg]{txfonts}
\usepackage{hyperref}

\newcommand{\COBOLD}{{\tt CO$^{\tt 5}$BOLD}}
\newcommand{\cobold}{{\tt CO$^{\tt 5}$BOLD}}
\newcommand{\LHD}{{\tt LHD}}
\newcommand{\LINFOR}{{\tt Linfor3D}}
\newcommand{\ATLAS}{{\tt ATLAS9}}
\newcommand{\MULTI}{{\tt MULTI}}
\newcommand{\BALMER}{{\tt BALMER}}
\newcommand{\MARCS}{{\tt MARCS}}
\newcommand{\teff}{\ensuremath{T_{\mathrm{eff}}}}
\newcommand{\logg}{\ensuremath{\mathrm{log}\,g}}
\newcommand{\MoH}{\ensuremath{\left[\mathrm{M}/\mathrm{H}\right]}}
\newcommand{\FeH}{\ensuremath{\left[\mathrm{Fe}/\mathrm{H}\right]}}
\newcommand{\aoFe}{\ensuremath{\left[\mathrm{\alpha}/\mathrm{Fe}\right]}}

\newcommand{\mlp}{\ensuremath{\alpha_{\mathrm{MLT}}}}
\newcommand{\xtmean}[1]{\ensuremath{\left\langle #1\right\rangle}}

\begin{document}

\title{Abundances of lithium, oxygen, and sodium in the turn-off stars of Galactic globular cluster 47~Tuc}

\author{
       V. Dobrovolskas
       \inst{1,2}
       \and
       A. Ku\v{c}inskas
       \inst{2}
       \and
       P.~Bonifacio
       \inst{3}
       \and
       S.~A.~Korotin
       \inst{4}
       \and
       M.~Steffen
       \inst{5,3}
       \and
       L.~Sbordone
       \inst{6,3}
       \and
       E.~Caffau
       \inst{6,3}
       \and
       H.-G.~Ludwig
       \inst{6,3}
       \and
       F.~Royer\inst{3}
       \and
       D.~Prakapavi\v{c}ius
       \inst{2}
       }

 \institute {Vilnius University Astronomical Observatory, M. K. \v{C}iurlionio 29, Vilnius, LT-03100, Lithuania
            \\
            \email{vidas.dobrovolskas@ff.vu.lt}
            \and
            Institute of Theoretical Physics and Astronomy, Vilnius University, Go\v{s}tauto 12, Vilnius, LT-01108, Lithuania
            \and
            GEPI, Observatoire de Paris, CNRS, Universit\'{e} Paris Diderot, Place Jules Janssen, 92190 Meudon, France
            \and
            Department of Astronomy and Astronomical Observatory, Odessa National University and Isaac Newton Institute of Chile Odessa branch, Shevchenko Park, 65014 Odessa, Ukraine
            \and
            Leibniz-Institut f\"ur Astrophysik Potsdam, An der Sternwarte 16, D-14482 Potsdam, Germany
            \and
            Landessternwarte -- Zentrum f\"ur Astronomie der Universit\"at Heidelberg, K\"{o}nigstuhl 12, D-69117 Heidelberg, Germany\\
            }

\date{Received ; accepted }

\abstract
  {The cluster 47 Tuc is among the most metal-rich Galactic globular clusters and its metallicity is similar to that of metal-poor disc stars and open clusters. Like other globular clusters, it displays variations in the abundances of elements lighter than Si, which is generally interpreted as evidence of the presence of multiple stellar populations.}
  {We aim to determine abundances of Li, O, and Na in a sample of of  110 turn-off (TO) stars, in order to study the evolution of light elements in this cluster and to put our results in perspective with observations of other globular and open clusters, as well as with field stars.
  }
  {We use medium resolution spectra obtained with the GIRAFFE spectrograph at the ESO 8.2m Kueyen VLT telescope and use state of the art 1D model atmospheres and NLTE line transfer to determine the abundances. We also employ \cobold\ hydrodynamical simulations to assess the impact of stellar granulation on the line formation and inferred abundances.
  }
  {Our results confirm the existence of Na-O abundance anti-correlation and hint towards a possible Li-O anti-correlation in the TO stars of 47~Tuc. At the same time, we find no convincing evidence supporting the existence of Li-Na correlation. The obtained 3D~NLTE mean lithium abundance in a sample of 94 TO stars where Li lines were detected reliably, $\langle A({\rm Li})_{\rm 3D~NLTE}\rangle = 1.78 \pm 0.18$\,dex, appears to be significantly lower than what is observed in other globular clusters. At the same time, star-to-star spread in Li abundance is also larger than seen in other clusters. The highest Li abundance observed in 47~Tuc is about 0.1\,dex lower than the lowest Li abundance observed among the un-depleted stars of the metal-poor open cluster NGC\,2243.
  }
  {The correlations/anti-correlations among light element abundances confirm that
  chemical enrichment history of 47~Tuc was similar to that of other globular clusters, despite the higher metallicity of 47~Tuc. The lithium abundances in 47~Tuc, when put into context with observations in other clusters and field stars, suggest that stars that are more metal-rich than $\FeH\sim -1.0$ experience significant lithium depletion during their lifetime on the main sequence, while the more metal-poor stars do not. Rather strikingly, our results suggest that initial lithium abundance with which the star was created may only depend on its age (the younger the star, the higher its Li content) and not on its metallicity.}

\authorrunning{Dobrovolskas et al.}
\titlerunning{Li, O, and Na abundances in 47~Tuc}

\maketitle

\section{Introduction}

Galactic globular clusters are important tracers of early Galactic chemical evolution. Although they have been long thought to be coeval and chemically homogeneous stellar populations, this view has changed dramatically during the last decade, driven by the spectroscopic \citep{GBB01,GSC04,MVP08,MMP09,GCB12} and photometric \citep{LJS99,BPA04,DBC05,PVB05,PBA07,MBP08,APK09,DHS09,HLJ09,KAM10a,KAM10b,MPK10,KAM11,PMA12} discoveries of chemical inhomogeneities which should not be present in simple stellar populations. It is commonly accepted today that globular clusters are made of at least two stellar populations that are characterised by different abundances in proton capture and alpha-elements, however, with little variation in the iron-peak and neutron capture elements. Moreover, star-to-star variations in the abundances of light elements are significantly larger than observational errors, besides, there are clear indications of various (anti-)correlations between the abundances of different light elements. All these findings may point to changes in chemical composition of the globular cluster stars induced by different proton and $\alpha$-particle capture reactions.

Surface (observed) chemical composition of red giant stars may be altered by mixing processes that may bring nuclear products from the deeper interiors to stellar surface. Such mixing processes, however, would not be able to account for the light element abundance correlations observed in unevolved stars (such as turn-off point stars and subgiants) since their core temperatures are not sufficiently high to start the NeNa or MgAl reaction cycles. Therefore, chemical abundance trends observed in unevolved globular cluster stars suggest that part of the cluster stars must have formed from gas already polluted by the previous generation stars. Although the nature of these primordial polluters is still under discussion, the most viable candidates are thought to be either intermediate mass Asymptotic Giant Branch stars \citep{VD05,DV07,dercole08,VD08,VD09,DDV10,DDC12} or fast rotating massive stars \citep{decressin07,DCS09,DBC10,CCD13}.

The reliability of stellar abundance estimates relies on the realism of the assumptions involved in the procedure of their determination. The overwhelming majority of abundance studies so far have utilised hydrostatic, one-dimensional (1D) stellar model atmospheres. Current three-dimensional (3D) hydrodynamical model atmospheres, on the other hand, are readily able to account for the non-stationary and multi-dimensional phenomena in stellar atmospheres (such as convection, granulation, wave activity, and so forth). Besides, there is a growing amount of evidence which shows that differences between the abundances derived using 3D hydrodynamical and classical 1D model atmospheres may indeed be important, reaching in their extremes to $-1.0$\,dex in case of atomic and to $-1.5$\,dex in case of molecular species at metallicities below $\MoH=-2.0$ \citep[][]{CAT07,KSL13,DKS13}. It is not surprising therefore that the number of studies using 3D hydrodynamical model atmospheres in stellar abundance work has been increasing very fast, especially during the last decade \citep[see, for example,][]{ANT00,CLS08,CLS10,GHB09,RAK09,CAN09,BBL10,DKA12}.

In this study we apply 3D hydrodynamical model atmospheres to investigate the evolution of light chemical elements lithium, oxygen, and sodium in a sample of 110 main-sequence turn-off (TO) stars of Galactic globular cluster 47~Tucanae. As a first step in this procedure, we determine the abundances of O and Na using classical \ATLAS\ 1D model atmospheres and 1D~NLTE spectral line synthesis performed with the \MULTI\ spectral synthesis package. We then utilise 3D hydrodynamical model atmospheres calculated with the \COBOLD\ model atmosphere code \citep[][]{FSL12} to estimate the 3D--1D abundance corrections and apply them to correct the 1D~NLTE abundances for 3D hydrodynamical effects. On the other hand, the 3D~NLTE abundance of lithium was determined using interpolation formula from \citet{SBC10} that was obtained using 3D hydrodynamic model atmospheres and NLTE spectral line synthesis. Our goal is therefore to obtain a reliable and homogeneous set of 3D~NLTE abundances of Li, O, and Na for a statistically large number of TO stars, which could be used further to study chemical enrichment history of 47~Tuc and to constrain possible chemical evolution scenarios of Galactic globular clusters.

The paper is structured as follows. In Sect.~\ref{sect:obs-data} we describe the observational data used in this study. The details of atmospheric parameter determination are outlined in Sect.~\ref{sect:atm-par}. Model atmospheres are described in Sect.~\ref{sect:atm-mod}. The procedure of 1D~LTE and 1D~NLTE abundance determinations and the impact of 3D hydrodynamical effects on the spectral line formation, as well as the resulting 3D--1D abundance corrections for Li, O, and Na, are described in Sect.~\ref{abundan}. Finally, we discuss the possible chemical evolution scenarios of 47~Tuc in Sect.~\ref{sect:discussion} and provide a short summary and conclusions in Sect.~\ref{sect:conclusions}.

\section{Observational data\label{sect:obs-data}}

Spectra of the TO stars in 47~Tuc investigated in this work were obtained with the GIRAFFE spectrograph in August--September, 2008, under the programme 081.D-0287(A) (PI: Shen). The same data set was independently analysed by \citet{DLG10}.

All program stars were observed in the Medusa mode, using three setups: HR15N (wavelength range $647.0-679.0$\,nm, resolution $R=17\,000$), HR18 ($746.8-788.9$\,nm, $R=18\,400$), and HR20A ($807.3-863.2$\,nm, $R=16\,036$). In each setup, 114--116 fibres were dedicated to the program stars and 14--16 were used for the sky spectra. In total, 12 exposures were obtained using HR15N setup, 10 using HR18, and 6 exposures using HR20A setup, each individual exposure lasting for 3600~s.

Raw spectra were bias-subtracted, flat-fielded and wavelength calibrated using the command-line version (v.~2.8.9) of GIRAFFE pipeline\footnote{\url{http://www.eso.org/sci/software/pipelines/giraffe/giraf-pipe-recipes.html}}. All sky spectra from each setup were median averaged and the obtained master sky spectrum was subtracted from each individual stellar spectrum using a custom-written IDL routine (there is no sky subtraction routine included in the standard pipeline package). After the sky subtraction, individual star spectra were corrected for the barycentric radial velocity and co-added to increase the signal-to-noise ratio. Signal-to-noise ratio of the final combined spectra was $S/N=80-90$ in the vicinity of oxygen triplet at $\lambda = 777$~nm. In total, spectra of 113 TO stars were extracted during the data reduction procedure. Finally, continuum normalisation of the GIRAFFE spectra was performed with the IRAF\footnote{IRAF is distributed by the National Optical Astronomy Observatories, which are operated by the Association of Universities for Research in Astronomy, Inc., under cooperative agreement with the National Science Foundation.} task \textit{continuum}.

\subsection{Cluster membership}

The initial stellar sample selection was based on the colour-magnitude diagram of 47~Tuc (Fig.~\ref{fig:cmd}). We then determined radial velocity of all selected stars to check whether they fulfill kinematic membership requirements. Radial velocity was calculated from the measured central wavelengths of two sodium lines (see Table~\ref{tab:atompar}), by taking the average value of the two measurements (line profile fit quality was not considered in the radial velocity determination procedure). Radial velocity from both sodium lines agreed to within 1.0 km\,s$^{-1}$ for 74 stars (66\,\% of the sample) while the largest difference on velocity values determined from the two lines never exceeded 3.5\,km\,s$^{-1}$.

Average barycentric radial velocity of the 113 sample stars is $-17.6$~km\,s$^{-1}$, which agrees well with the value of $-18.0$~km\,s$^{-1}$ listed for this cluster in the catalogue of \citet{H96}. Our obtained radial velocity dispersion is 7.2~km\,s$^{-1}$, with the lowest and highest velocity values of $-32.0$~km\,s$^{-1}$ and +1.3~km\,s$^{-1}$, respectively. We note that the velocity dispersion determined by us is slightly smaller than that obtained by \citet{H96}, 11.0~km\,s$^{-1}$. This, however, should be expected since all stars studied here are located between 4\farcm5 and 11\farcm5 from the cluster centre, that is, beyond its half-light radius of 3\farcm17 \citep{H96}. Our radial velocity results lead us to conclude that all selected stars are highly probable members of 47~Tuc.

\section{Atmospheric parameters\label{sect:atm-par}}

\subsection{Effective temperatures\label{sect:teff}}

Effective temperature of the program stars was determined by fitting wings of the observed H$\alpha$ line profiles with the theoretical H$\alpha$ profiles \citep{FAG93, barklem08, CVA11}. Theoretical line profiles were computed in LTE with a modified version of the Kurucz's \BALMER\ code\footnote{The original version of the code is available from \url{http://kurucz.harvard.edu}}, which allowed the usage of different theories for the self-broadening and Stark broadening of the line profile \citep[see][]{SBC10}. In our case, we used the theory of \citet{barklem00a,barklem00b} for the self-broadening and that of \citet{stehle} for the Stark broadening. A grid of input models for computing synthetic H$\alpha$ profiles was calculated using \ATLAS\ model atmosphere code (see Sect. \ref{sect:atm-mod}
for details).
In the fitting procedure, we only used H$\alpha$ line wings ($\geqq$90\% of the normalised flux) which are most temperature sensitive. We avoided the line core region because it forms in the outer atmospheric layers where deviations from the LTE are possible and, thus, the LTE approach used in the \BALMER\ code may not be adequate (Fig.~\ref{fig:halpha}). We also excluded two spectral lines located on the wings of the H$\alpha$ profile, to avoid possible systematic shifts in the effective temperature determination. In Fig.~\ref{fig:halpha} we show the spectral region around H$\alpha$ line and highlight the regions that were used in (and excluded from) the fitting procedure.

\begin{figure}[tb]
\centering
\includegraphics[width=\columnwidth]{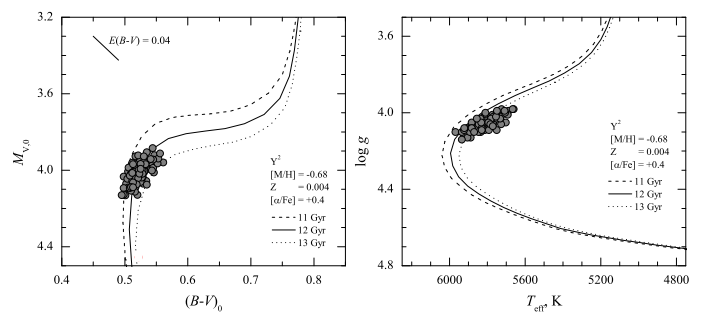}
 \caption{
 \textbf{Left:} de-reddened $M_{\rm V,0}-(B-V)_0$ CMD of the TO stars studied in 47~Tuc. Lines are Yonsei-Yale (Y$^2$) isochrones (11, 12, and 13~Gyr), computed assuming $Z=0.004$ ($\MoH=-0.68$) and $\aoFe=+0.4$. Line in the upper left corner shows the reddening vector assuming $A_{\rm V}/E(B-V)=3.1$.
 \textbf{Right:} HR diagram showing the TO sample stars in 47~Tuc, with the effective temperatures and gravities determined as described in Sect.~\ref{sect:teff} and \ref{sect:logg}. Solid lines are Y$^2$ isochrones.}
\label{fig:cmd}
\end{figure}

\begin{figure}[tb]
 \centering
 \includegraphics[width=\columnwidth]{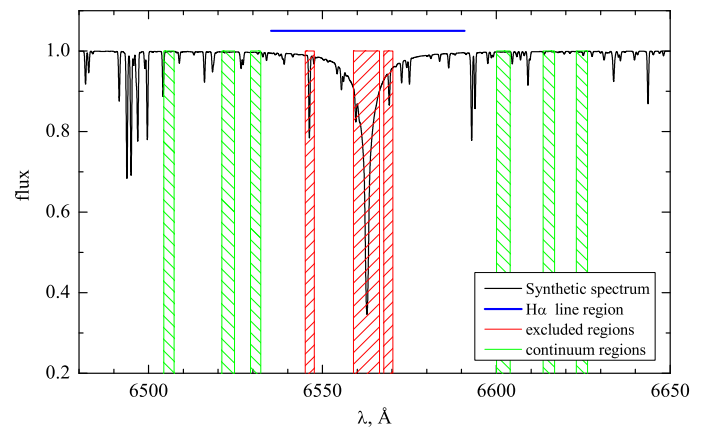}
 \caption{Synthetic spectrum of a TO star in 47~Tuc (00006129, $\teff=5851$~K, $\log g=4.06$) showing H$\alpha$ line region used for the effective temperature determination. Blue bar above the synthetic spectrum indicates the range were fitting of the H$\alpha$ line wings was done, red dashed rectangles mark the regions excluded from the fitting procedure, while green dashed rectangles highlight the spectral regions used to determine continuum level.}
 \label{fig:halpha}
\end{figure}

It is important to note that effective temperatures determined from the H$\alpha$ line wings are sensitive to surface gravity \citep[see, e.g.][]{SBC10}. On the other hand, surface gravities of the studied TO stars were derived using empirical formula that requires the knowledge of effective temperatures (Sect.~\ref{sect:logg}). Therefore, effective temperatures and gravities of the TO stars studied here were derived using iterative procedure, by adjusting \teff\ and $\log g$ simultaneously. However, because the program stars occupy very narrow range both in \teff\ and $\log g$ (see Fig.~\ref{fig:cmd}, Table~\ref{tab:Tuc47-atmpar}), the corrections applied during the iterative procedure typically did not exceed $\pm10$~K and $\pm0.05$~dex, respectively.

We note that there is a systematic difference in the effective temperatures derived from the H$\alpha$ line wings and those that would be inferred from the Yonsei-Yale isochrone, with differences becoming larger at lower \teff\ (Fig.~\ref{fig:cmd}). For the majority of stars, however, this difference does not exceed $\sim120$~K. In terms of abundances, this would translate to differences of $\sim0.1$~dex in case of lithium and oxygen, and $\sim0.08$~dex in case of sodium (see Sect.~\ref{sect:abn-sens}), and would have a minor influence on the results obtained in this work.

\subsection{Surface gravities\label{sect:logg}}

Iron ionisation equilibrium condition enforcement is one of the most widely used methods to estimate stellar surface gravity. However, we were not able to use this approach due to the small number -- only two -- of ionised iron lines available in our spectra. Instead, we determined surface gravities using the following relation (with \teff\ derived from the H$\alpha$ line wings)
\begin{equation}
 \logg\ = 4.44 + \mathrm{log}(M/M_\odot) - \mathrm{log}(L/L_\odot) + 4\mathrm{log}\teff - 4\mathrm{log}\teff^{\odot}\, ,
\label{eqn:logg}
\end{equation}
\noindent where $\teff^{\odot}=5780$~K is the adopted solar effective temperature, \textit{M} and \textit{L} are stellar mass and luminosity, respectively (sub-/upper-script $\odot$ denotes solar values). Stellar luminosity of the individual stars was determined from the Yonsei-Yale isochrones (see Sect.~\ref{subsect:mass}) using their absolute magnitudes, $M_{\rm V}$, derived from photometry.

In Sect.~\ref{subsect:mv}--\ref{subsect:mass} below we outline the procedures used to derive absolute magnitudes and masses, i.e. the quantities needed to obtain surface gravities of the program stars using Eq.~\ref{eqn:logg}.

\subsubsection{Absolute magnitudes $M_V$\label{subsect:mv}}

We used $BV$ photometry from \citet{BS09} to determine $M_V$ of our program stars. During the visual inspection, we have found two possible photometric blends among the sample of program stars, namely stars with designations 00043029 ($V_1=17.83$~mag, $V_2=18.49$~mag, separation = 1\farcs5) and  00112168 ($V_1=17.64$~mag, $V_2=18.65$~mag, separation = 0\farcs5). These stars were unresolved or marginally resolved by the GIRAFFE fibre having 1\farcs2 aperture on the sky, which led to incorrect determination of their atmospheric parameters. The two stars were therefore excluded from the abundance analysis.

Because of its high Galactic latitude \citep[$b=-45$\degr,][]{H96}, 47~Tuc experiences little interstellar reddening: the values found in the literature range from $E(B-V)=0.032$ (\citet{SFD98}, \footnote{\url{http://irsa.ipac.caltech.edu/applications/DUST/}}) to $E(B-V)=0.055$ \citep{GFC97}. Similarly, \citet{GSA02} obtained $E(B-V)=0.04$ from the Str\"{o}mgren $uvby$ photometry. We note that the reddening uncertainty of $\Delta E(B-V)=0.02$ would lead to the uncertainty of $\Delta\log g\approx0.02$ in the surface gravity and therefore would have a minor impact on the effective temperature determination. To correct for the interstellar reddening, we adopted $E(B-V)=0.04$ from \citet{H96}. The distance modulus, $V-M_{\rm V}=13.37$, was also taken from \citet{H96}.

\subsubsection{Stellar mass\label{subsect:mass}}

We determined stellar mass from the comparison of the observed colour-magnitude diagram (CMD) of 47~Tuc with theoretical isochrones. For this purpose, we used $t=11$, 12, and 13\,Gyr age Yonsei-Yale\footnote{\url{http://www.astro.yale.edu/demarque/yyiso.html}} (Y$^2$) isochrones, computed for the metallicity of Z = 0.004 and $\alpha-$element enhancement of $\aoFe=+0.4$. The observed CMD was fitted best with the 12--13 Gyr Y$^2$ isochrones (Fig.~\ref{fig:cmd}). The narrow magnitude range occupied by the program stars ($V=17.225-17.507$~mag) translates into a mass interval of $\sim0.02$\,M$_{\odot}$ (Table~\ref{tab:isochrones}). Using Eq.~\ref{eqn:logg} we find that the uncertainty in stellar mass of $\Delta M= 0.02$\,M$_\odot$ leads to a change in surface gravity of only $\Delta\log g\approx0.01$ (while keeping the effective temperature and luminosity fixed).

\begin{table}
 \begin{center}
\caption{Masses of the program stars derived from the Y$^2$ isochrones of different age.
\label{tab:isochrones}}
  \begin{tabular}{ccc}
  \hline\hline
\noalign{\smallskip}
 \textit{t}, Gyr & mass range, \textit{M}$_\odot$ & $\langle M \rangle$, \textit{M}$_\odot$ \\
  \hline\noalign{\smallskip}
   12      & 0.84 - 0.86         & 0.85                \\
   13      & 0.82 - 0.84         & 0.83                \\
  \hline
  \end{tabular}
  \end{center}
\end{table}

Small mass range of the program stars, together with a small change in the average mass between the 12 and 13~Gyr isochrones, led us to assume a fixed average mass of 0.84~M$_\odot$ for all program stars.

\subsection{Final sample}

The final stellar sample used in this work contained 110 TO stars. Their atmospheric parameters derived using the prescriptions given in the previous sections are provided in Table~\ref{tab:Tuc47-atmpar}.

\section{Model atmospheres\label{sect:atm-mod}}

The following model atmospheres were used in our study:

\begin{itemize}
\item
\textbf{\ATLAS:} these 1D hydrostatic stellar model atmospheres were computed using the \ATLAS\ code \citep{Kur93, SBC04, S05}, with new opacity distribution functions as described in \citet{CK03} for alpha-element enhanced ($\aoFe=+0.4$) chemical composition and microturbulence velocity of 1.0~km\,s$^{-1}$. \ATLAS\ model atmospheres used in the 1D~NLTE analysis of oxygen and sodium abundances with the \MULTI\ spectral synthesis code were calculated with a mixing length parameter $\mlp=1.25$. In addition, \ATLAS\ model atmospheres used in the effective temperature determination were calculated with a mixing length parameter $\mlp=0.5$, according to the recommendations of \citet{FAG93} and \citet[][]{VV96}; we note though that the choice of $\mlp$ had a negligible influence on the derived effective temperatures;
\item
\LHD: these 1D hydrostatic model atmospheres were calculated using the \LHD\ model atmosphere code. The code utilises the same equation of state and opacities as those used with the 3D hydrodynamical \COBOLD\ model atmospheres (see below), thus allowing a differential comparison of the 3D and 1D predictions to be made \citep[for more about the \LHD\ model atmosphere code see, e.g.,][]{CL07}. In this study, \LHD\ 1D hydrostatic model atmospheres were used together with the 3D hydrodynamical \COBOLD\ model atmospheres to compute the 3D--1D abundance corrections for oxygen and sodium.
\item
\COBOLD: 3D hydrodynamical model atmospheres used in our work were calculated with the \COBOLD\ model atmosphere code \citep{FSL12}, which solves time-dependent equations of hydrodynamics and radiation transfer on a Cartesian grid. The \COBOLD\ models were used to assess the influence of convection on the thermal structures of stellar atmospheres and spectral line formation, and to compute the 3D--1D abundance corrections for oxygen and sodium.
\end{itemize}

\section{Determination of elemental abundances\label{abundan}}

Abundances of lithium, oxygen, and sodium in the TO stars of 47~Tuc were determined using slightly different procedures. For sodium and oxygen, the 1D~NLTE abundances were derived by best-fitting the observed line profiles with the synthetic spectra computed with the 1D~NLTE spectral synthesis code \MULTI. For this purpose, we slightly expanded and modified the model atoms of oxygen and sodium used with the \MULTI\ code. The obtained 1D~NLTE abundances of oxygen and sodium were then corrected for the 3D effects using 3D--1D abundance corrections computed with the 3D hydrodynamical and 1D hydrostatic model atmosphere codes \COBOLD\ and \LHD, respectively.

The abundance of lithium, on the other hand, was determined using the equivalent widths of lithium lines measured in the spectra of the program stars. Then, we used the interpolation formula from \citet[][]{SBC10} to directly obtain the 3D~NLTE lithium abundance estimate.

In the following sections we will focus on the steps involved in the abundance determinations of all three elements discussed here.

\subsection{Spectral lines and their atomic parameters}

Atomic parameters of spectral lines used in the abundance derivations of lithium, oxygen, and sodium are provided in Table~\ref{tab:atompar}. Central line (777.416~nm) of the oxygen triplet was significantly affected by telluric emission and thus proved to be unsuitable for the abundance determinations. Oxygen abundance was therefore derived using the two remaining lines of the oxygen triplet located at 777.194~nm and 777.539~nm.

\begin{table}[tb]
\setlength{\tabcolsep}{3pt}
\begin{center}
\caption{Atomic parameters of the spectral lines used in the abundance determinations of lithium, oxygen, and sodium.
\label{tab:atompar}}
\begin{tabular}{llcrllc}
   \hline\hline
   \noalign{\smallskip}
Element & $\lambda$, nm  & $\chi$, eV & log \textit{gf}$^{\mathrm{a}}$ & log $\gamma_{rad}$$^{\mathrm{b}}$ & log $\frac{\gamma_4}{N_e}$$^{\mathrm{c}}$ & log $\frac{\gamma_6}{N_H}$$^{\mathrm{d}}$ \\
  \hline\noalign{\smallskip}
\ion{Li}{i} & 670.776      & 0.000    & $-0.009$          & 7.56    & $-5.78$ & $-7.574$ \\
\ion{Li}{i} & 670.791      & 0.000    & $-0.309$          & 7.56    & $-5.78$ & $-7.574$ \\
\ion{O}{i}  & 777.194      & 9.146    &   0.324          & 7.52     & $-5.55$ & $-7.443^{\mathrm{e}}$ \\
\ion{O}{i}  & 777.539      & 9.146    & $-0.046$          & 7.52    & $-5.55$ & $-7.443^{\mathrm{e}}$ \\
\ion{Na}{i} & 818.326      & 2.102    &   0.230          & 7.52$^{\mathrm{e}}$    &   $-5.62^{\mathrm{e}}$ & $-7.425^{\mathrm{e}}$ \\
\ion{Na}{i} & 819.482      & 2.104    &   0.490          & 7.52$^{\mathrm{e}}$    &   $-5.62^{\mathrm{e}}$ & $-7.425^{\mathrm{e}}$ \\
   \hline
\end{tabular}
\end{center}
\begin{list}{}{}
\item[$^{\mathrm{a}}$] \citet{Kur93}; $^{\mathrm{b}}$ natural broadening constant \citep[][]{KRP00}; $^{\mathrm{c}}$ Stark broadening constant  \citep[][]{KRP00}; $^{\mathrm{d}}$ van der Waals broadening constant \citep[][]{KRP00}; $^{\mathrm{e}}$ classical value \citep[][]{C05b}.
\end{list}
\end{table}

\begin{figure*}[tbh]
\centering
\includegraphics[width=17cm]{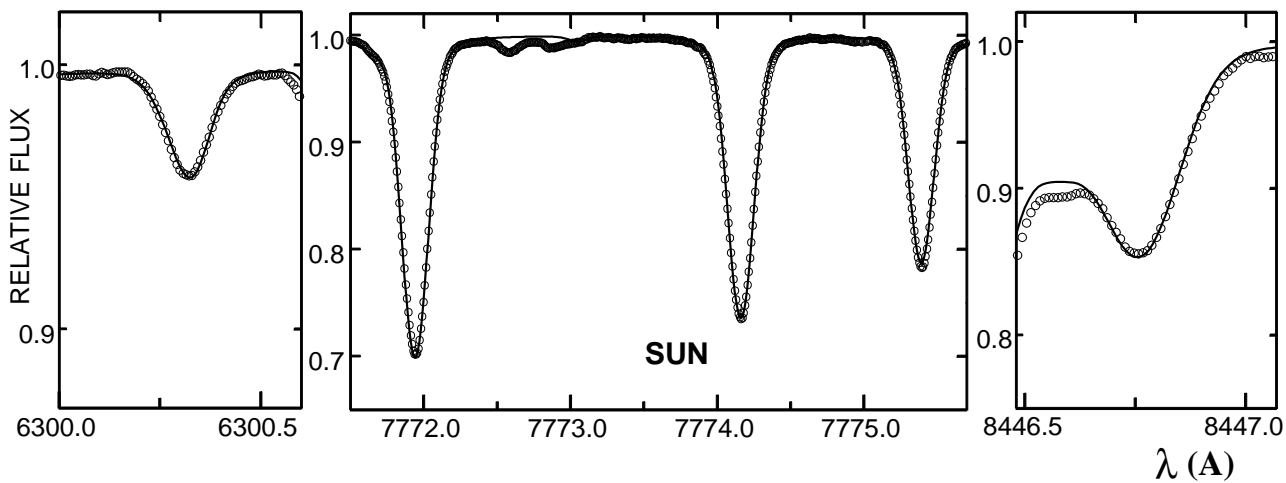}
 \caption{Synthetic NLTE spectral lines of oxygen (solid lines) compared with the observed solar spectrum (dots) from \citet{KFB84}. \textbf{Left}: forbidden line at 630.0~nm, \textbf{Centre}: infrared triplet at 777~nm, \textbf{Right}: infrared line at 845\,nm. All synthetic spectral line profiles were computed using solar oxygen abundance of $A{\rm (O)} = 8.71$.}
\label{fig:sunOnlte}
\end{figure*}

\subsection{3D+NLTE abundances of oxygen and sodium}

The 1D~NLTE abundances of oxygen and sodium were derived by fitting the observed spectral line profiles with the synthetic spectra computed using the 1D spectral synthesis code \MULTI. We then corrected the obtained 1D~NLTE abundances for the 3D hydrodynamical effects, by using 3D--1D abundance corrections calculated with the 3D hydrodynamical \COBOLD\ and 1D \LHD\ model atmospheres. In the latter two cases, spectral line synthesis was done using the \LINFOR\ package. A constant microturbulence velocity of $\xi_{\rm micro}=1.0$~km\,s$^{-1}$ was assumed for all program stars in the spectral synthesis calculations using the 1D model atmospheres (i.e., \ATLAS\ and \LHD), irrespective whether it was done with the \MULTI\ or \LINFOR\ line synthesis packages. We note though, that none of the spectral lines analyzed in this paper were strongly saturated and therefore the resulting abundances were insensitive to the choice of microturbulence velocity (see Table \ref{tab:abnsens}). The steps involved in the derivation of oxygen and sodium abundances are summarised below.

\begin{figure*}[tb]
\centering
\includegraphics[width=17cm]{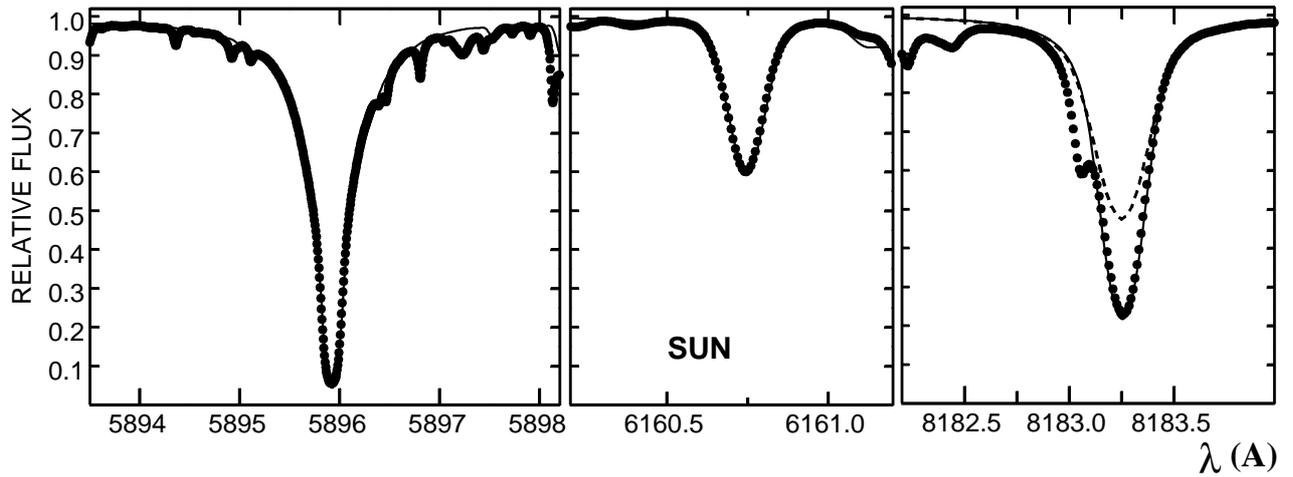}
 \caption{Synthetic 1D~NLTE spectral lines of sodium (solid lines) compared with the observed solar flux spectrum (dots) from \citet{KFB84}. In case of 818~nm line, which experiences the strongest deviations from the LTE, we also show the LTE line profile (dashed line). All synthetic spectral lines were computed using solar sodium abundance of $A{\rm (Na)} = 6.25$.}
\label{fig:sunNanlte}
\end{figure*}

\begin{figure*}[tb]
\centering
\includegraphics[width=17cm]{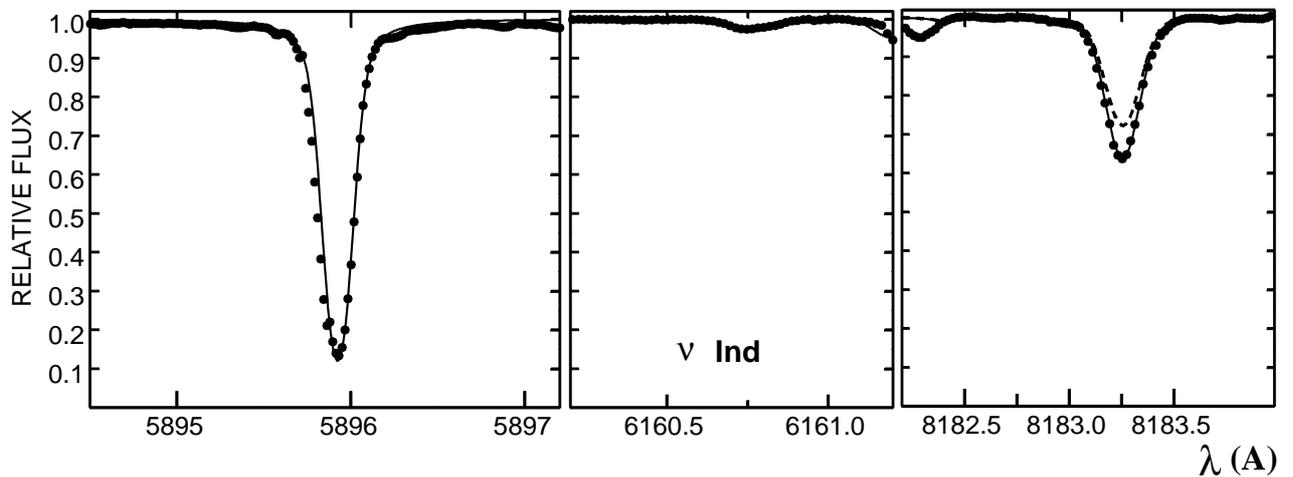}
 \caption{Synthetic 1D~NLTE spectral lines of sodium (solid lines) compared with the observed spectrum of $\nu$~Ind spectrum \citep[dots, taken from the UVES archive of the Paranal Observatory Project][]{BJL03}. All synthetic spectral line profiles were computed using sodium abundance of $A{\rm (Na)} = 4.5$.}
\label{fig:nuIndNanlte}
\end{figure*}

\subsubsection{Model model atom of oxygen}

As a first step in the 1D~NLTE abundance analysis, we modified and expanded the model atoms of oxygen and sodium. In the case of oxygen, we started with the model atom described in \citet{MKK00} and updated it with new collisional transition rates. The updated version of the model atom currently consists of 23 energy levels of \ion{O}{i} and the ground level of \ion{O}{ii} for which detailed statistical equilibrium equations are solved. Additional 48 energy levels of \ion{O}{i} and 15 energy levels of the higher ionisation stages were included to account for the particle number conservation. All 46 bound-bound transitions with the wavelengths shorter than 10\,000~nm were used in the calculation of atomic level population numbers. Ionisation cross-sections were taken from the TOPBASE \citep{CMO93}. Rate coefficients for the bound-bound electronic collisional transitions  obtained using detailed quantum mechanical calculations by \citet{barklem07} were used with the lowest seven energy levels of \ion{O}{i}. Other allowed transitions were approximated by the classical formula of \citet{regemorter}, while for the forbidden transitions the formula of \citet{allen73} was used, with $\Omega = 1$. In terms of its complexity, the updated model atom therefore occupies an intermediate position between its older version presented in \citet{MKK00} and a model atom utilised in, e.g., \citet{FAB09} and/or \citet{SMR13}. We note, however, that the Solar oxygen abundance obtained with the updated model atom is still very similar to that derived either by \citet{MKK00} or \citet{SMR13} (see this Section below).

Oscillator strengths of the oxygen lines at 630.0 and 636.3~nm were taken from \citet{SZ00}, while for the rest of the lines we used data from NIST. We utilised van der Waals line broadening constants obtained using quantum mechanical calculations \citep{AO95,BO97}. One should note that at higher metallicities the \ion{Ni}{i} line located at 630.0~nm becomes an important contributor to the strength of the forbidden oxygen line at 630.0\,nm. To account for the nickel blend, we used $\log gf=-2.11$ \citep{JLL07} for the \ion{Ni}{i} 630.0~nm line and a nickel abundance of $A{\rm (Ni)}=6.17$ \citep{SAG09} in the spectral synthesis calculations. Isotopic splitting into two components with $\lambda$($^{58}{\rm Ni})=630.0335$~nm and $\lambda$($^{60}{\rm Ni})=630.0355$~nm \citep{BFL04} was taken into account as well.

Since collisions with hydrogen atoms play an important role in the atmospheres of cool stars, this effect was taken into account by using the classical formula of \citet{drawin}, in the form suggested by \citet{SH84} and with a correction factor $S_{\rm H}=1/3$. The numerical value of this coefficient was chosen by comparing predicted and observed oxygen line profiles in the solar spectrum. In particular, by setting $S_{\rm H}=0.0$ it was not possible to reconcile the observed strengths of the IR triplet lines at 777.1-5~nm and the line at 844.6~nm, as this line and those belonging to the triplet system would imply different solar oxygen abundance. On the other hand, when the coefficient was set to 1.0 then the IR triplet was too weak and did not produce the same abundance estimate as that obtained from the forbidden oxygen line at 630.0~nm.

To test the updated model atom, we compared synthetic spectral line profiles with those observed in the spectrum of the Sun. We used the solar model atmosphere from \citet{CK03}, together with the chromosphere model VAL-C and the corresponding microturbulence velocity distribution from \citet{VAL81}. In addition to testing the realism of the model atom, this also allowed us to investigate the possible influence of the chromosphere on the NLTE line formation. Test calculations showed, however, that the latter effect was very small: the difference in the equivalent widths of IR oxygen triplet lines computed using the model atmospheres with and without the chromosphere was less than 1.5\%. To compare the theoretical lines profiles with those observed in the spectrum of the Sun, the re-reduced \citep{Kur06} Kitt Peak Solar Flux Atlas \citep{KFB84} was used (R = 523\,000, signal-to-noise ratio S/N = 4000 at the wavelength of the infrared sodium lines). In addition, we compared the synthetic line profiles computed for the centre of the solar disc with the observed ones from the atlas of \citet{DNR73}. Synthetic lines were convolved with a Gaussian profile, to obtain the same spectral resolution as the Kitt Peak Solar Flux Atlas, and broadened by 1.8~km\,s$^{-1}$ rotational velocity. We assumed $\xi_{\rm micro}$ = 1.0~km\,s$^{-1}$ microturbulence velocity and $\zeta_{\rm macro}$ = 2.0~km\,s$^{-1}$ macroturbulence velocity. Solar oxygen abundance derived from the IR triplet lines was $A{\rm (O)}\footnote{Abundance \textit{A} of element X, \textit{A}(X,) is defined as $A{\rm (X)} = \log \epsilon{\rm (X)} = \log (N_{\rm X}/N_{\rm H}) + 12$, where $N_{\rm X}$ and $N_{\rm H}$ are number densities of element X and hydrogen, respectively.} = 8.71$, which agrees well with $A{\rm (O)} =  8.71$ obtained by \citet{SAG09} and $A{\rm (O)} = 8.69$ recommended by \citet{AGS09}. It is also in good agreement with $A{\rm (O)} = 8.73$ obtained by \citet{MKK00} and $A{\rm (O)} = 8.75$ derived by \citet{SMR13}, with the abundances in both cases measured using IR triplet lines. Comparison of the calculated and observed oxygen line profiles in the solar spectrum is shown in Fig.~\ref{fig:sunOnlte}.

\subsubsection{Model atom of sodium}

In the case of sodium, we started with the model atom constructed by \citet[][]{KM99}. The updated sodium model atom currently consists of 20 energy levels of \ion{Na}{i} and the ground level of \ion{Na}{ii}. Fine splitting has been taken into account only for the 3p level, in order to ensure reliable calculations of the sodium doublet transitions at 589~nm. In total, 46 radiative transitions were taken into account for the level population number calculations. Fixed radiative transition rates were used for other weak transitions. Photoionisation cross-sections were taken from the TOPBASE database \citep{CMO93}.

The sodium model atom was updated to accommodate new rate coefficients of collisions with electrons and hydrogen atoms and ionisation by the hydrogen atoms for the lower 9 levels of \ion{Na}{i}, which were obtained by \citet{BBD10} using quantum mechanical computations. For other levels, the classical formula of Drawin has been used in the form suggested by \citet{SH84}, with the correction factor $S_{\rm H}=1/3$. One should note however, that the $S_{\rm H}$ parameter plays a minor role here because the spectral lines of sodium studied in this work arise from the transitions between the lower energy levels, and for these transitions we used the rate coefficients of collisions with hydrogen from \citet{BBD10} (i.e., no $S_{\rm H}$ correction factor was applied for these levels). On the other hand, collisions with hydrogen have little influence on the change in the population numbers of levels with high-excitation energy and, therefore, do not influence the population numbers of the lowest 9 energy levels. The highest sensitivity to $S_{\rm H}$ factor is seen in the case of sodium lines located at 568.2 and 568.8 nm. However, in case of the Sun, the change in $S_{\rm H}$ from 0.0 to 1.0 leads to the change in the equivalent widths of these particular lines of only $\approx1\%$. The change in the equivalent widths of other spectral lines is less than one percent.

Rate coefficients for the collisions with electrons have been revised, too. In this case, we used  electron collision cross-sections from \citet{ISB08} for transitions between the lowest eight energy levels of the sodium atom, for a wide range of impacting electron energies. Electron ionisation cross-sections were also taken from \citet{ISB08}. For the rest of the allowed transitions we used the relation of \citet{regemorter}, while for the forbidden transitions the formula of \citet{allen73} was utilised.

\begin{figure}[tb]
\centering
\includegraphics[width=\columnwidth]{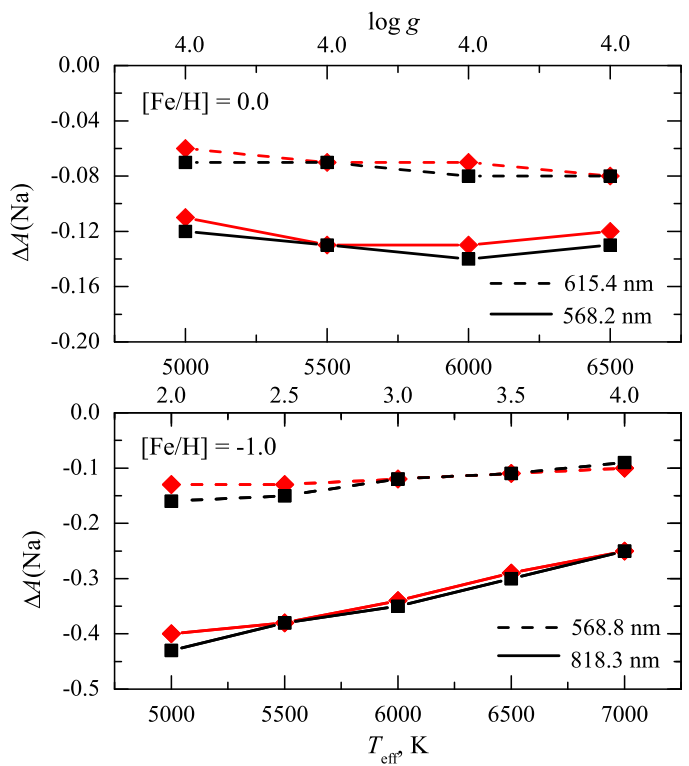}
 \caption{1D~NLTE abundance corrections for different spectral lines of sodium, as obtained in this work (black lines and symbols) and by \citet[][red lines and symbols]{Lind11}. \textbf{Top} panel: abundance corrections for the spectral lines located at $\lambda = 568.2$ nm (solid lines) and $\lambda = 615.4$ nm (dashed lines). \textbf{Bottom} panel: abundance corrections for the spectral lines located at $\lambda = 568.8$ nm (dashed lines) and $\lambda = 818.3$ nm (solid lines).}
\label{fig:Na_delta_nlte}
\end{figure}

To test the modified model atom, we carried out calculations of the spectral line profiles for the Sun, Procyon, and $\nu$~Ind. For the Sun, we used re-reduced Kitt Peak Solar Flux Atlas \citep{Kur06}. In case of Procyon and $\nu$~Ind, spectra were taken from the UVES archive of the Paranal Observatory Project \citep{BJL03}. The latter spectra were obtained with the UVES spectrograph and spectral resolution of R = 80\,000, and have a S/N of more than 300. In case of the Sun, the synthetic sodium lines were convolved with a Gaussian profile to obtain the spectral resolution of the Kitt Peak Solar Atlas, and then further rotationally broadened by 1.8~km\,s$^{-1}$. Microturbulence and macroturbulence velocities were set to $\xi_{\rm micro}$ = 1.0~km\,s$^{-1}$ and $\zeta_{\rm macro}$ = 2.0~km\,s$^{-1}$, respectively. The average solar sodium abundance determined from nine sodium lines (located at 514.88, 568.26, 568.82, 588.99, 589.59, 615.42, 616.08, 818.33, and 819.48~nm) is $A{\rm (Na)} (= \log \epsilon{\rm (Na)})$ = $6.25 \pm 0.08$\,dex which agrees well both with the solar photospheric abundance of $A{\rm (Na)}=6.24\pm0.04$~dex from \citet{AGS09} and with currently recommended solar abundance of $A{\rm (Na)}=6.29\pm0.04$~dex from \citet{lodders09}.

If the model atom is constructed correctly it must describe adequately the spectral lines belonging to different multiplets, yielding identical abundance of a given chemical element irrespective to which line of the multiplet is used. To perform such test, we have chosen lines with different sensitivities to NLTE effects. For example, sodium lines at 818.3 and 819.4~nm are very sensitive to NLTE effects while the widely-used lines at 615.4 and 616.0~nm are not affected by strong deviations from the LTE. In Fig.~\ref{fig:sunNanlte} we show the observed spectrum of the Sun, together with the synthetic spectral line profiles computed under the assumption of NLTE. Clearly, synthetic line profiles fit well the observed lines belonging to different multiplets. For comparison, we also show the LTE line profile of the line located at 818.3~nm which is amongst the most sensitive to NLTE effects. In the case of metal-poor stars, the sodium lines at 615.4 and 616.0~nm are too weak to be measured reliably, thus the sodium abundance has to be determined using stronger lines. To further test the realism of the modified sodium model atom we therefore synthesised several stronger lines of sodium in the spectrum of $\nu$~Ind (HD~211998; $\teff=5240$~K, $\log g=3.43$, $\FeH=-1.6$). We conclude that also in this case all synthetic lines computed with a single sodium abundance of $A{\rm (Na)} = 4.5$ fit the observed spectrum satisfactorily (Fig.~\ref{fig:nuIndNanlte}).

In addition, we compared 1D~NLTE abundance corrections derived for the sodium lines using our model atom with those obtained by \citet{Lind11}. The sodium model atoms used in the two studies are very similar. In particular, collisions with hydrogen atoms in both cases are accounted for by using quantum mechanical computation data from \citet{BBD10}. 1D~NLTE abundance corrections obtained with our model atom are in good agreement with the results of \citet{Lind11} over a wide range of stellar atmospheric parameters (Fig.~\ref{fig:Na_delta_nlte}).

\subsubsection{1D~NLTE abundances of O and Na}

The newly updated model atoms of oxygen and sodium described in the previous sections were  used to derive abundances of the two elements in the TO stars of 47~Tuc. For this purpose, we employed the spectral synthesis code \MULTI\ \citep{carlsson} in its modified version \citep{KAL99}, while the abundances were determined by fitting synthetic line profiles to those in the observed spectra of TO stars using $\chi^2$ minimisation. Throughout the spectral synthesis computations we used fixed microturbulence $\xi_{\rm micro} = 1.0$~km\,s$^{-1}$. Macroturbulence velocity was varied as a free parameter to achieve the best fit to the observed line profiles, with its typical values determined in the range of 1--5~km\,s$^{-1}$.

\begin{figure*}[tbh]
\centering
\includegraphics{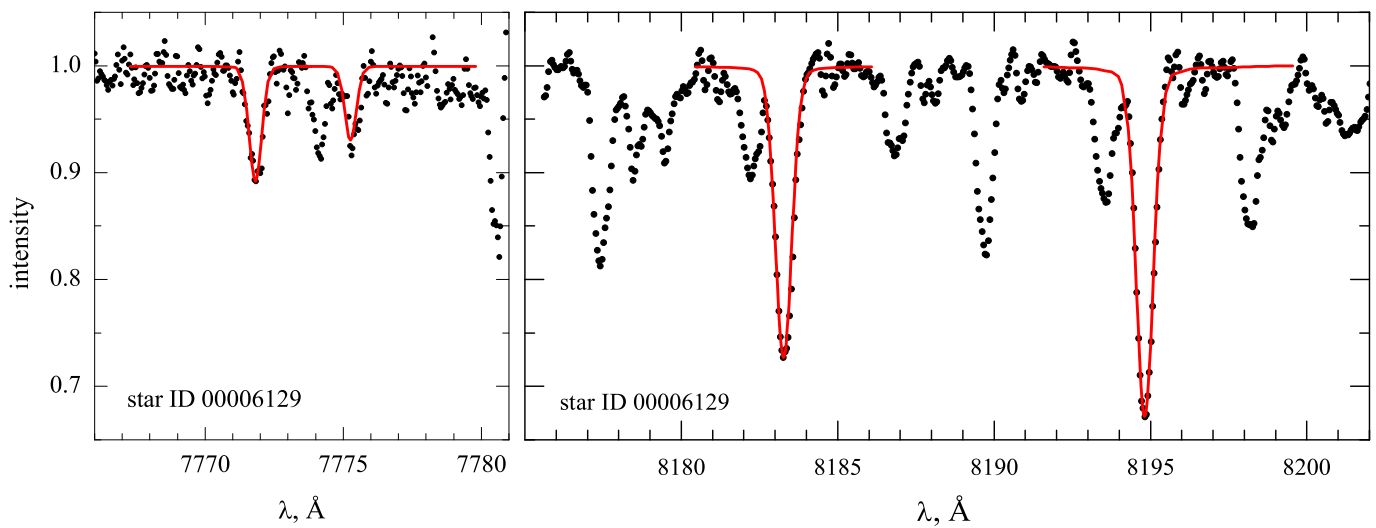}
 \caption{Typical observed GIRAFFE spectrum of TO star in 47~Tuc (star ID 00006129, see Table~\ref{tab:Tuc47-atmpar}, dotted lines), together with synthetic spectrum (red solid lines) computed using the \MULTI\ code and fitted to the oxygen 777~nm triplet (\textbf{left} panel), sodium 818.3~nm (\textbf{centre} panel), and sodium 819.5~nm (\textbf{right} panel) lines.}
\label{fig:spect}
\end{figure*}

Examples of the observed and best-fitted synthetic 1D~NLTE spectral line profiles are shown in Fig.~\ref{fig:spect}, while the determined 1D~NLTE abundances of O and Na are provided in Table~\ref{tab:Tuc47-atmpar} (Appendix~\ref{sect:AppA}). We note that typical differences between the abundances of oxygen and sodium obtained using 1D~NLTE and LTE spectral line synthesis (estimated by fitting the NLTE and LTE synthetic profiles to the observed line profile) are indeed significant, $\Delta_{\rm 1D\,NLTE-LTE}\approx-0.20$ and $\approx-0.35$~dex, respectively.

\begin{table}[tb]
\caption{Parameters of the 3D hydrodynamical \COBOLD\ atmosphere models used in this work.\label{tab:3Dmodels2}}   
\centering                 
\setlength{\tabcolsep}{3pt}
\begin{tabular}{c c c c c}     
\hline\hline                   
  \teff, K & \logg\ & \MoH\ & Grid dimension, Mm & resolution\\    
                   &          &       & x $\times$ y $\times$ z & x $\times$ y $\times$ z\\
\hline                     
 5475 & 4.0 &   0    & $20.3\times20.3\times10.6$ & $140\times140\times150$ \\    
 5533 & 4.0 & $-1$   & $19.9\times19.9\times10.6$ & $140\times140\times150$ \\
 5927 & 4.0 &  0     & $25.8\times25.8\times12.5$ & $140\times140\times150$ \\
 5850 & 4.0 & $-1$   & $25.8\times25.8\times12.5$ & $140\times140\times150$ \\
\hline                     
\end{tabular}
\end{table}

\subsubsection{3D--1D abundance corrections for O and Na\label{sect:3Dabund}}

Convection has a significant impact on the spectral line formation in the atmospheres of cool stars. A number of recent studies have shown that treating convection in one-dimensional hydrostatic model atmospheres with mixing-length theory may lead to significant differences in the abundances of chemical elements with respect to those determined using 3D hydrodynamical model atmospheres
\citep[a non-exhaustive list includes][]{asp99,CAT07,CL07,jonay,BCu,DKA12}. The role of convection becomes especially important in the atmospheres of metal-poor stars ($\FeH<-2$), where horizontal fluctuations of thermodynamic quantities and changes in the vertical temperature and velocity profiles may lead to significant differences in the predicted spectral line strengths. We therefore used 3D hydrodynamic \COBOLD\ stellar model atmospheres \citep{FSL12} to assess the impact of such effects on the spectral line formation in the atmospheres of our program stars.

The \COBOLD\ models used for this purpose were taken from the CIFIST 3D hydrodynamical model atmosphere grid \citep{LCS09}. Since the model spacing in the $\teff-\log g-\MoH$ plane is rather coarse, there are no CIFIST models with the atmospheric parameters exactly corresponding to those of the program stars. We therefore used four \COBOLD\ models bracketing the parameters of TO stars with their \teff\ and \MoH. The desired quantities (e.g., line strengths) were computed using each of the four models and then interpolated to the effective temperature and metallicity of a given TO star. Atmospheric parameters of the 3D hydrodynamical models are provided in Table~\ref{tab:3Dmodels2}. Each simulation run covered about $\approx7.5$\,days in stellar time, or $\approx$19 convective turnover times as measured by the Brunt-Vais\"{a}l\"{a} timescale \citep[see][for the definition; we note that the advection timescale is always significantly shorter]{KSL13}. Monochromatic opacities used in the model calculations were taken from the \MARCS\ stellar atmosphere package \citep{GEK08} and were grouped into a number of opacity bins using the opacity binning technique \citep{N82,L92,LJS94,VBS04}. Five opacity bins were used for the $\MoH=0.0$ models and six bins for the $\MoH=-1.0$ models. The models were computed using solar-scaled elemental abundances from \citet{AGS05}, by applying a constant enhancement in the alpha-element abundances of $\aoFe=+0.4$ for the models at $\MoH=-1.0$. All model simulations were performed under the assumption of local thermodynamic equilibrium, LTE, with scattering treated as true absorption \citep[for more details on the model calculations see][]{LCS09}.

To perform spectral line synthesis calculations, from the four 3D hydrodynamical model runs we selected four smaller subsamples of 20 model structures (snapshots; 18 were selected in the case of model at $\teff=5927$~K, $\log g=4.0$, $\MoH=0.$). The selected snapshots were spaced nearly equidistantly in time and spanned the entire length of each simulation run. Each snapshot ensemble was selected in such a way as to ensure that its most important statistical properties (the average effective temperature, its standard deviation, mean velocity at optical depth unity, mean velocity and residual mass flux profiles) would match those of the entire simulation run as closely as possible. Time separation between the individual snapshots in the 20 snapshot ensemble was $\approx0.4$~days ($\approx1$ convective turnover time) which allows us to consider them statistically independent.

As in our previous work \citep[e.g.,][]{KSL13,DKS13}, the influence of convection on the spectral line formation was assessed with the help of 3D--1D abundance corrections. This correction, $\Delta_{\rm 3D - 1D}$, is defined as a difference between the abundance $A{\rm (X)}$ of chemical element \textit{X} derived at the observed equivalent width of a given spectral line using 3D and 1D model atmospheres, $\Delta_{\rm 3D - 1D} = A(X)_{\rm 3D} - A(X)_{\rm 1D}$. We also made use of two additional abundance corrections, $\Delta_{\rm 3D-\langle3D\rangle}\equiv A({\rm X_{i}})_{\rm 3D}-A({\rm X_{i}})_{\rm \langle3D\rangle}$, and $\Delta_{\rm \langle3D\rangle-1D}\equiv A({\rm X_{i}})_{\rm \langle3D\rangle}-A({\rm  X_{i}})_{\rm 1D}$. These corrections utilise average $\xtmean{\mbox{3D}}$ models which were computed by horizontally averaging all atmospheric structures in the twenty 3D model snapshot ensemble (the fourth power of temperature was averaged on surfaces of equal optical depth). Obviously, the average $\xtmean{\mbox{3D}}$ models do not contain information about the horizontal fluctuations of dynamical and thermodynamic quantities. Therefore, the first of the two corrections, $\Delta_{\rm 3D-\langle3D\rangle}$, allows the importance of the horizontal fluctuations to be assessed, while the other, $\Delta_{\rm \langle3D\rangle-1D}$, the role of differences between the temperature profiles of the average $\xtmean{\mbox{3D}}$ and 1D models. The total 3D--1D abundance correction is indeed the sum of the two constituents, $\Delta_{\rm 3D-1D}\equiv\Delta_{\rm 3D-\langle3D\rangle}+\Delta_{\rm \langle3D\rangle-1D}$.

To compute the abundance corrections, 3D hydrodynamical, average $\xtmean{\mbox{3D}}$, and 1D \LHD\ model atmospheres were used to synthesise spectral lines with the equivalent widths, $EW$, equal to those measured in the given program star, and to obtain the 3D, $\xtmean{\mbox{3D}}$, and 1D abundances of a given chemical element \citep[note that this is different from what was done in][where only very weak lines were used to compute abundance corrections]{DKS13,KSL13}. The resulting $\Delta_{\rm 3D-\langle3D\rangle}$, $\Delta_{\rm \langle3D\rangle-1D}$, and $\Delta_{\rm 3D-1D}$ abundance corrections were then interpolated to the effective temperature of the given program star and the metallicity of 47~Tuc (the latter was kept fixed at $\FeH=-0.7$). We note that the \COBOLD\ and \LHD\ model atmospheres were computed using the same atmospheric parameters, equation of state, and opacities, in order to minimise the possible sources of discrepancies in their predicted line strengths. This allowed us to focus solely on the differences arising due to different treatment of convection in the 3D hydrodynamical and 1D hydrostatic model atmospheres.

We thus computed 3D--1D abundance corrections for oxygen and sodium, for every object in the sample of 110 TO stars studied here and for every spectral line used in the abundance determination, using the line equivalent widths obtained during the 1D~NLTE abundance analysis. The 1D~NLTE abundances were then corrected for the 3D effects, by adding the average 3D--1D abundance correction obtained for a given element in a given star to its average 1D~NLTE abundance. Abundances obtained using such procedure will be hereafter referred to as 3D+NLTE abundances, in order to make a clear distinction from the 3D~NLTE abundances of lithium which were obtained based on full 3D~NLTE spectral line synthesis calculations. The obtained 3D+NLTE abundances of oxygen and sodium are provided in Table~\ref{tab:Tuc47-atmpar} (Appendix~\ref{sect:AppA}).

\begin{table*}[tb]
 \caption{Average, minimum, and maximum abundance corrections for the spectral lines of Li, O, and Na in the sample of 110 TO stars in 47~Tuc.\label{tab:AbuCorr}}
 \centering
 \begin{tabular}{cc|ccc|ccc|ccc}
 \hline\hline
 species & $\lambda$,\,nm  &      & $\Delta_{\rm 3D-<3D>}$  &  &   & $\Delta_{\rm <3D> - 1D}$ &   &  & $\Delta_{\rm 3D - 1D}$   &    \\
         &                 &   aver  &  min    & max     &  aver   & min     & max    &   aver   & min     & max    \\
 \hline
 \ion{Li}{i}    & 670.8           & $-0.156$ & $-0.157$  & $-0.155$ &  +0.047 & +0.031  & +0.062 &  $-0.110$ & $-0.125$ & $-0.095$\\
 \ion{O}{i}     & 777.2           &   0.000 & $-0.028$  & +0.015  &  +0.067 & +0.049  & +0.074 &  +0.067  & +0.021  & +0.089 \\
 \ion{O}{i}     & 777.5           & $-0.012$ & $-0.033$  & +0.007  &  +0.061 & +0.046  & +0.072 &  +0.049  & +0.013  & +0.079 \\
 \ion{Na}{i}    & 818.3           & $-0.063$ & $-0.098$  & $-0.043$ &  +0.066 & +0.044  & +0.093 &  +0.003  & $-0.037$ & +0.031 \\
 \ion{Na}{i}    & 819.5           & $-0.056$ & $-0.076$  & $-0.041$ &  +0.075 & +0.048  & +0.102 &  +0.019  & $-0.013$ & +0.042 \\ \hline
 \end{tabular}
\end{table*}

The information about the obtained abundance corrections is summarised in Table~\ref{tab:AbuCorr}, where we provide $\Delta_{\rm 3D-\langle3D\rangle}$, $\Delta_{\rm \langle3D\rangle-1D}$, and $\Delta_{\rm 3D-1D}$ abundance corrections for the spectral lines of oxygen and sodium used in our study. In each case, we list the average, minimum and maximum values of the correction computed from the ensemble of individual corrections corresponding to each of 110 TO stars in 47~Tuc. Obviously, the 3D--1D abundance corrections are small and typically do not exceed $\Delta_{\rm 3D-1D}\approx0.06$\,dex.

Finally, we would like to warn the reader that the procedure used by us to obtain 3D+NLTE abundances should be utilised with caution. The reason for this is that population numbers of atomic levels in NLTE depend very sensitively on temperature but this is not taken into account by applying 3D--1D LTE corrections to 1D~NLTE abundances. Our test simulations utilising full 3D~NLTE radiation transfer and  the 3D hydrodynamical models used above show that in the case of lithium the 3D+NLTE approach may in fact be justifiable at solar metallicity (Appendix~\ref{sect:Li-test}). However, at $\MoH=-1.0$ (and below) the full $\Delta_{\rm 3D\,NLTE-1D~\,LTE}$ abundance correction becomes significantly different from the combined $\Delta_{\rm 1D\,NLTE-LTE} + \Delta_{\rm 3D-1D}$ correction. Such deviation occurs because the lithium line formation extends rather far into the outer atmosphere where the amplitude of horizontal temperature fluctuations (and thus, its influence on the atomic level populations) is largest. Moreover, weaker line blanketing in the metal-poor stellar atmospheres leads to more efficient photoionisation. All this may result in significantly different population numbers in 3D~NLTE and 1D~NLTE cases. Clearly, these differences cannot be accounted for in the 3D+NLTE approach, by applying 3D--1D corrections to the 1D~NLTE abundances. In this respect, the situation is somewhat safer with oxygen and sodium since in these two cases the contribution of horizontal temperature fluctuations and differences between the average 3D and 1D temperature profiles to the total 3D--1D abundance correction are about equal but of opposite sign, which leads to smaller total abundance correction, $\Delta_{\rm 3D-1D}$ (see Table~\ref{tab:AbuCorr}). It is nevertheless obvious that full 3D~NLTE spectral synthesis should be utilised whenever such possibility is available; hopefully, this may gradually become accessible with the implementation of NLTE methodology into the 3D spectral synthesis codes \citep{LMA13,PSK13,HS13}.

\subsection{3D~NLTE abundances of lithium}

As a first step in the abundance analysis of lithium, we determined the equivalent width of the lithium 670.8~nm resonance doublet by fitting the observed spectrum of a given star with a synthetic line profile computed using 1D~NLTE \texttt{SYNTHE} spectral synthesis package \citep{Kur05,S05}. We then used the obtained equivalent widths to determine 3D~NLTE lithium abundance by using analytical formula (B.1) from \citet{SBC10}. This interpolation formula was obtained by utilising the results of 3D~NLTE spectral synthesis computations done for a range of lithium abundances and by covering the effective temperatures and surface gravities typical to those of the main sequence stars. The fitting formula of \citet{SBC10} was derived using models in the metallicity range of $\FeH=-1.0$ to --3.0, thus for the stars in 47 Tuc we were extrapolating to slightly higher metallicities. 

We were able to detect the lithium resonance doublet in 94 TO stars. The derived 3D NLTE lithium abundances span the range of $1.24 < A({\rm Li}) < 2.21$\,dex, with the average value of $\langle A({\rm Li})\rangle = 1.78\pm0.18$ (the error is standard deviation of lithium abundance in the ensemble of 94 TO stars). The obtained lithium abundances are provided in Table~\ref{tab:Tuc47-atmpar} (Appendix~\ref{sect:AppA}).

\begin{table}[tb]
 \begin{center}
 \caption{Li, O, and Na abundance sensitivity to changes in the atmospheric parameters.
 \label{tab:abnsens}}
 \begin{tabular}{lcccc}
 \hline\hline
 \noalign{\smallskip}
 Element & $\Delta\,T_{\rm eff}$ & $\Delta\,\rm{log}\,g$ & $\Delta\,\xi_{\rm micro}$ & $\Delta A$ \\
         & $\pm100$\,K           & $\pm0.1$\,dex         & $\pm0.2\,\rm{km\,s^{-1}}$ & dex        \\
 \hline\noalign{\smallskip}
 \ion{Li}{i} & +0.09   & $-0.01$ &   0.00  & 0.09  \\
             & $-0.08$ & +0.01   &   0.00  & 0.08  \\
 \ion{O}{i}  & $-0.08$ & +0.03   & $-0.01$ & 0.09  \\
             & +0.09   & $-0.03$ &  +0.01  & 0.10  \\
 \ion{Na}{i} & +0.06   & $-0.05$ & $-0.03$ & 0.08  \\
             & $-0.07$ & +0.04   &  +0.02  & 0.08  \\
 \hline
 \end{tabular}
 \end{center}
\end{table}

\subsection{Abundances sensitivity to changes in the atmospheric parameters \label{sect:abn-sens}}

The influence of the uncertainties in the atmospheric parameters to the abundance determinations of Li, O and Na was assessed by varying atmospheric parameters within their typical uncertainties: effective temperature by $\pm100$~K, surface gravity by $\pm0.1$~dex, and microturbulence velocity by $\pm0.2$~km\,s$^{-1}$. We took the average atmospheric parameters and spectral line strengths measured in the TO stars of 47~Tuc as reference values for this test. The corresponding changes in the elemental abundances are provided in Table~\ref{tab:abnsens}. The numbers in case of lithium were obtained by varying the corresponding atmospheric parameters in the formula of \citet{SBC10}, while for oxygen and sodium sensitivity determination was made using \ATLAS\ model atmospheres and 1D~NLTE line synthesis with \MULTI. The last column contains all three abundance changes added in quadrature and thus may serve as a measure of combined sensitivity to changes in the uncertainty all atmospheric parameters. The results show that the uncertainty in the effective temperature has by far the largest impact on the abundance determination of all three elements investigated in this work.

\subsection{Sensitivity of sodium abundances to blending with CN lines}

Spectral region around the sodium lines used in this study contains several weak CN lines which may blend with the lines of sodium. Since TO stars in 47~Tuc stars show large spread in carbon and nitrogen abundances \citep{CCB98, CGL05}, CN blends with sodium lines may introduce systematic changes in the derived sodium abundances, and may thus distort the resulting sodium abundance correlations. We therefore deemed it necessary to test the impact of CN spectral lines on the sodium abundance determination.

To this end, we synthesised a number of synthetic spectra with the Linux version \citep{S05} of the spectral synthesis code \texttt{SYNTHE} \citep{Kur05}, by using a number of \ATLAS\ model atmospheres that corresponded to the average and extreme values of atmospheric parameters of the studied TO stars in 47~Tuc. The spectra were synthesised using different combinations of C and N abundances representing the most CN-rich and CN-rich stars, as well as those with the average CN abundance (see Table~\ref{tab:CNabn}). As a reference, we also computed synthetic spectrum neglecting CN lines. Equivalent widths of synthetic sodium lines were then measured in each synthetic spectrum and sodium abundance was determined using the measured equivalent width and Linux version of the \texttt{WIDTH} code \citep{Kur93,Kur05,C05a}. We found that difference in the sodium abundance obtained from the spectrum without CN lines and that computed with the nitrogen enhancement of [N/Fe] = +1.25~dex is $\approx0.02$~dex for the sodium 819~nm line and $<0.01$~dex for the 818~nm line. We therefore conclude that the impact of CN line blending on the sodium abundance derivations may be safely ignored.

\begin{table}[tb]
 \begin{center}
 \caption{Combinations of carbon and nitrogen abundances used to estimate the impact of blending with CN lines on the determined sodium abundances.
 \label{tab:CNabn}}
  \begin{tabular}{cc}
  \hline\hline
  \noalign{\smallskip}
  [C/Fe] & [N/Fe] \\
  \hline\noalign{\smallskip}
  $-0.10$ & $-0.35$\\
  $-0.25$ &  +0.50 \\
  $-0.45$ &  +1.25 \\
  \hline
  \end{tabular}
 \end{center}
\end{table}

\section{Discussion\label{sect:discussion}}

{\bf
\subsection{Abundance anticorrelations}
}

\begin{figure}[tb]
\centering
\includegraphics[width=8.5cm]{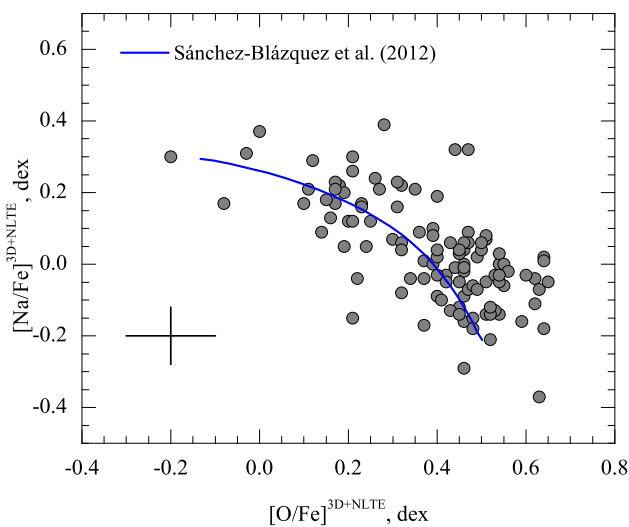}
 \caption{Abundances of oxygen and sodium in the TO stars of 47~Tuc derived in this work by taking into account 3D hydrodynamical and NLTE effects. Blue solid line shows chemical evolution model of 47~Tuc from \citet{SMG12}.
 }
\label{fig:OFe_NaFe_abn}
\end{figure}

In Fig.~\ref{fig:OFe_NaFe_abn} we plot the determined abundances of oxygen against those of sodium. Our results confirm the presence of the Na--O abundance anti-correlation. This finding is in good agreement with the results of \citet{DLG10} although we obtain a slightly smaller spread in the sodium abundance (Fig.~\ref{fig:OFe_NaFe_abn}). It is interesting to note that the observed Na--O abundance anti-correlation agrees surprisingly well with the model predictions of \citet{SMG12}.

We find that the effects of convection play a minor role in the spectral line formation of O and Na taking place in the atmospheres of TO stars in 47~Tuc, leading to relatively small $\Delta_{\rm 3D-1D}$ abundance corrections, $-0.04\dots+0.04$~dex for sodium and $+0.01\dots+0.09$~dex for oxygen. On the other hand, deviations from LTE are substantial: on average, the NLTE abundance correction is as large as $-0.35$~dex for sodium and $-0.20$~dex for oxygen.

\begin{figure}[tb]
\centering
\includegraphics[width=8.5cm]{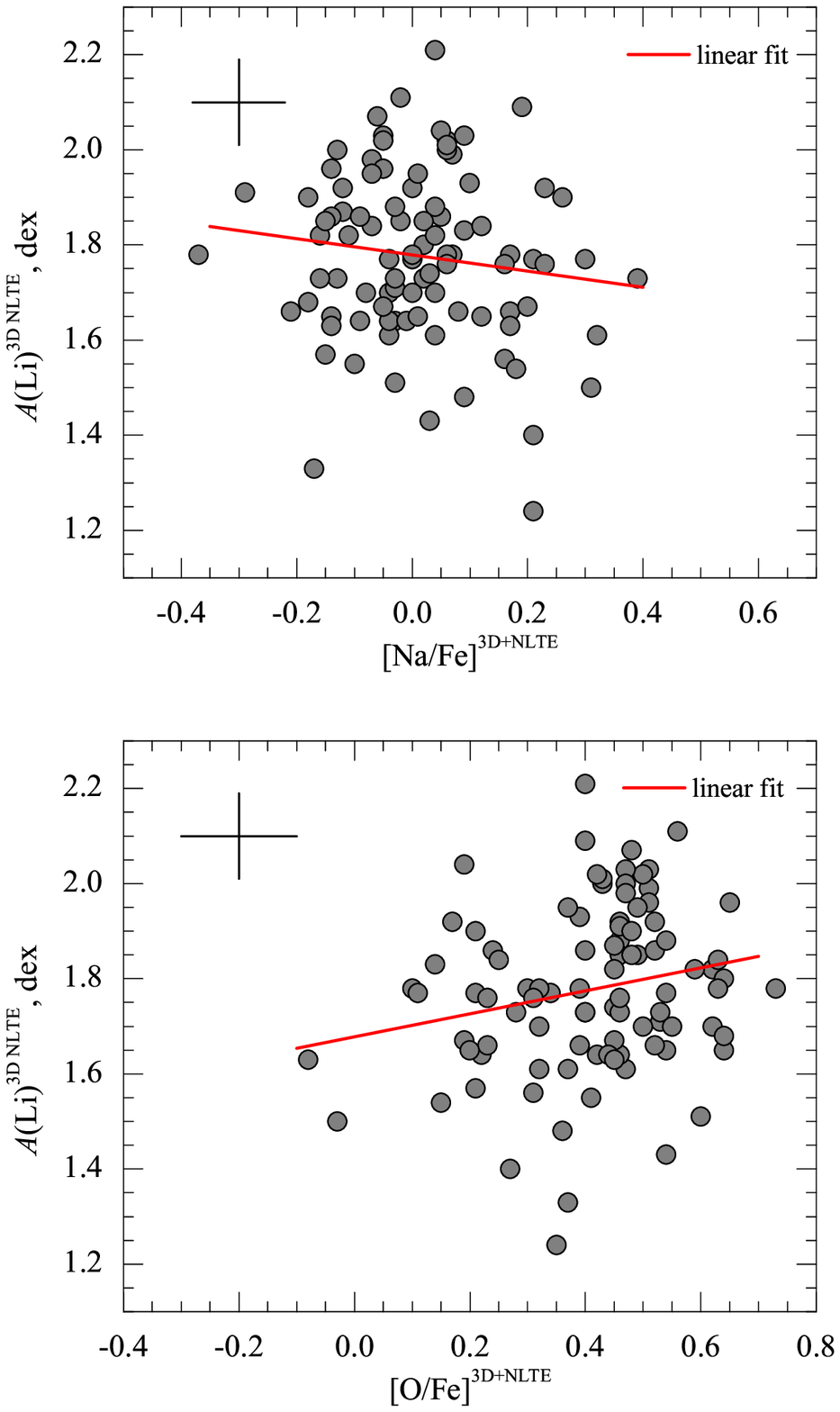}
 \caption{Lithium--sodium (\textbf{top} panel) and lithium--oxygen (\textbf{bottom} panel) abundances determined in this work. Unweighted linear fits to the data are shown by red solid lines.
 }
 \label{fig:LiONa_corr}
\end{figure}

\begin{table*}
\setlength{\tabcolsep}{4pt}
\caption{Average lithium abundance and its spread in Galactic globular
clusters and the open cluster NGC\,2243.}
\label{ligc}
\begin{center}
\begin{tabular}{lrccccl}
\hline
\\
Cluster       &   N  & [Fe/H] & $\langle A({\rm Li})\rangle$ & $A(Li)_{min}$ & $A(Li)_{max}$ & Reference       \\
\hline
\\
M\,4          &  73  & $-1.33$ &   2.14                       &    1.82       &    2.40       & \citet{Monaco12} \\
M\,92         &   6  & $-2.31$ &   2.35                       &    2.04       &    2.55       & \citet{B02}      \\
NGC\,6397     &  79  & $-2.02$ &   2.30                       &    2.01       &    2.52       & \cite{GHB09}     \\
NGC\,6752     & 102  & $-1.42$ &   2.29                       &    1.84       &    2.61       & \citet{Shen}     \\
$\omega$\,Cen &  52  & $-1.71$ &   2.19                       &    1.76       &    2.49       & \citet{Monaco10} \\
47\,Tuc       &  94  & $-0.76$ &   1.78                       &    1.24       &    2.21       & This paper       \\
\hline
\\
NGC\,2243     &  20  & $-0.55$ &   2.61                       &    2.33       &    2.85       & \citet{francois} \\
\\
\hline
\end{tabular}
\end{center}
\end{table*}

Although less convincingly, our results also hint towards the existence of Li--O correlation (Fig.~\ref{fig:LiONa_corr}). This is supported by the results of Kendall's tau ($\tau$) test \citep{Crecipes} which detects the existence of  Li--O correlation at 95\,\% probability level, with $\tau=0.14$. Although one should note that the data scatter is large, this result is nevertheless robust even if the two stars with the lowest oxygen abundance are excluded from the test. On the other hand, evidence of the Li--Na anti-correlation is weak: in this case, Kendall's tau test yields the detection at the level of only 58\,\% ($\tau=-0.06$).

\subsection{Evolution of lithium abundances inferred from data on star clusters.\label{clusters}}

Lithium, along with hydrogen and helium, was synthesised during the Big-Bang nucleosynthesis what makes it particularly important element because of its relevance to cosmology. On the assumption that its abundance in the oldest stars has not been altered since the star formation, the knowledge of the lithium abundance may allow the models of primordial nucleosynthesis to be tested. In the warm metal-poor stars the lithium abundance is roughly constant, \textit{A}(Li) = 2.1 -- 2.3, what is known as \textit{Spite Plateau} \citep{S82a,S82b,SBC10}. Primordial lithium abundance based on the WMAP measurements of the baryonic density \citep{SVP03,komatsu} and Standard Big Bang Nucleosynthesis (SBBN) is predicted to be \textit{A}(Li) = 2.7 \citep{cyburt}. The value derived from the measurements of Planck satellite \citep{planck13} is the same within errors \citep{CUV13}. It is still eludes explanation why the  \textit{Spite Plateau} is $\approx3$ times lower than the predicted primordial abundance \citep[see][for a discussion of possible explanations]{SBC10}.

The average  lithium abundance obtained in our study ($\langle A({\rm Li})\rangle = 1.78\pm0.18$) is in good agreement with the value of $A({\rm Li})=1.84 \pm 0.25$ obtained by \citet{BPM07} from the spectra of 4 TO stars in 47~Tuc and is lower than $\langle A({\rm Li})\rangle = 2.02 \pm 0.21$ obtained by \citet{DLG10} using the same set of spectra. Most likely, the difference between us and \citet{DLG10} is due to the different methods of analysis employed to derive lithium abundances.

\begin{figure}
\centering
\includegraphics[width=\columnwidth]{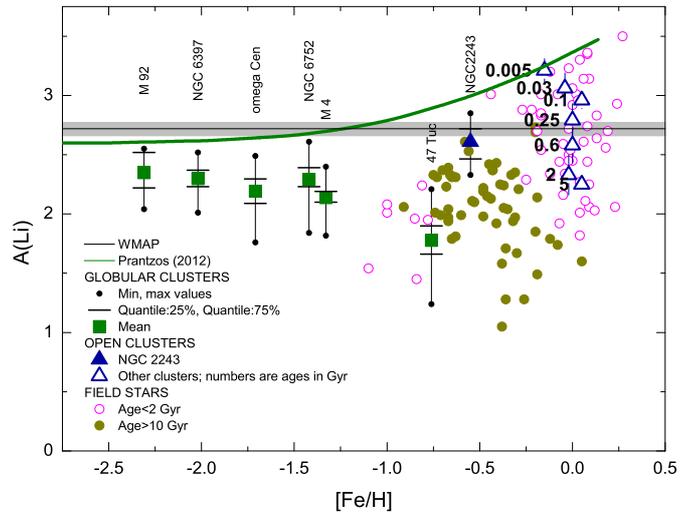}
 \caption{Lithium abundances in Galactic globular and open clusters, and field stars, plotted versus metallicity. We show the mean (filled green rectangles), minimum, and maximum values (small black dots connected by vertical solid lines), and 25\% and 75\% quantiles (black horizontal bars) of lithium abundances as derived in unevolved stars of Galactic globular clusters M\,92, NGC\,6397, $\omega$\,Cen, NGC\,6752, and 47 Tuc, and the old, metal-poor open cluster NGC\,2243 \citep[filled blue triangle, $\sim4.4$\,Gyr,][see also Table~\ref{ligc} for details]{Kaluzny}. The data for open clusters (open blue triangles) were taken from \citet[][]{SR05}, with each datapoint typically representing averages for several open clusters in a given age bin (ages are indicated by numbers next to the symbols, in Gyr, error bars in this case mark 1$\sigma$ spread in lithium abundances). Data for the Galactic field stars (filled and open circles) were taken from \citet{Lambert} and are shown for two age sub-groups: stars that are older than 10\,Gyr (filled circles) and stars that are younger than 2~Gyr (open circles). The solid green line shows theoretically predicted evolution of the Galactic lithium abundance from \citet{prantzos}.
}
\label{fig:plot_GC}
\end{figure}

The spread in the derived lithium abundances in 47~Tuc appears to be larger and the mean abundance lower than what is found in other globular clusters: if the exceptionally Li-rich stars Cl* NGC~6397 K~1657 \citep{Koch} and Cl* M~4 M~37934 \citep{Monaco12} are excluded, other globulars show a rather uniform lithium abundance. In Table~\ref{ligc} we have assembled literature data of the mean $A({\rm Li})$ and the range in $A({\rm Li})$ variation for the handful of globular clusters, and added to these the metal-poor open cluster NGC\,2243 \citep{francois}. When confronted with several analyses of the same cluster in the literature we chose, when available, the ones with the 3D~NLTE lithium abundances obtained using the fitting formula of \citet{SBC10}, to be directly comparable to the present results. For M~92 we chose the reanalysis of \citet{B02} rather than the original analysis of \citet{Boesgaard}, though this choice would bear no consequences to our discussion. The globular cluster NGC~6752 \citep{Shen} is the one with the wider range in Li abundances, after 47~Tuc, and has a higher mean $A({\rm Li})$ abundance, too. $\omega$ Cen is not an ordinary globular cluster but rather the stripped core of a satellite galaxy, and is the only cluster in Table~\ref{ligc} that shows a large spread in $\FeH$. In Table~\ref{ligc} and Fig.~\ref{fig:plot_GC} we adopted $\FeH=-1.71$ as a median of the metallicities in the \citet{Monaco10} sample. In Fig.~\ref{fig:plot_GC} we have also plotted the values for old (4.35 Gyr, \citealt{Kaluzny}), metal-poor ($\FeH=-0.54$, \citealt{francois}) open cluster NGC~2243, from the analysis of \citet{francois}. We also complemented the plot with the data for a number of open clusters having different ages \citep[taken from][]{SR05}, and Galactic field stars from \citet[][]{Lambert}.

There are several facts that are immediately evident from the inspection of Fig.~\ref{fig:plot_GC}:
\begin{itemize}
\item all globular clusters display a dispersion in lithium abundances, probably due to the chemical evolution of the cluster itself, as suggested by the Li--O correlation and, possibly, Li--Na anti-correlation;
\item this dispersion is small with respect to the ``gap'' between the mean cluster abundance and the primordial lithium abundance predicted by SBBN and WMAP measurements of the baryonic density;
\item the mean lithium abundance of the metal-poor globular clusters (i.e., all except 47~Tuc) traces well the \textit{Spite Plateau};
\item 47~Tuc has lower mean lithium abundance and higher dispersion than other globular clusters;
\item the mean lithium abundance and its dispersion in 47~Tuc are compatible with lithium abundances observed in the field stars at the same metallicity and older than 12~Gyr;
\item metal poor, old open cluster NGC~2243, with a metallicity close to that of 47~Tuc, has higher mean Li abundance;
\item field stars younger than 2~Gyr have, on average, higher lithium abundances than those that are older than 12~Gyr;
\item at approximately solar metallicity, there is a clear tendency for open clusters to have lower lithium abundance with increasing age.
\end{itemize}

These facts may be understood in terms of the following scenario. The stars in the globular clusters all form with the same Li abundance. This abundance may be slightly altered in the course of the star's life \citep[see][]{GHB09}. On top of this effect, one observes Li differences amongst the stars of first and successive generations, due to the effect of pollution of the cluster medium by the first generation stars (this would explain the Li--O correlation). All these effects are of second order, so that the mean cluster abundance is close to the original initial abundance, which is confirmed by nearly uniform Li abundance in all metal-poor clusters seen in Fig.~\ref{fig:plot_GC}. In the most metal-rich clusters, like 47~Tuc, the convective envelope is deeper and photospheric material is brought down to layers where the temperature is sufficient to destroy lithium. For this reason the mean Li abundance of 47~Tuc is lower and its Li dispersion is larger than in the other globular clusters. The old metal-rich field stars follow the same fate, and this is why, on average, they have a lower Li abundance. As time passes, Li is produced in the Galaxy so that the stars and clusters formed more recently are formed with a higher Li abundance. The open cluster NGC~2243 clearly shows this: in spite of the fact that its metallicity is only slightly slightly different from that of 47~Tuc, its mean Li abundance is clearly higher. Importantly, the highest Li abundance found in 47~Tuc is lower than the lowest abundance observed in NGC~2243 (excluding the Li-dip stars, of course). A prediction of this simplistic qualitative scenario, would be that, by analogy to what is observed in 47~Tuc, when  NGC~2243 will reach the age of 12~Gyr it should have a mean Li abundance by 0.5~dex lower than its present-day value.

Chemical evolution of lithium in the Galaxy is difficult to model, due to the existence of several sources that can, potentially, produce lithium besides the Big Bang Nucleosynthesis \citep[see][for a review]{Matteucci}. Current models assume that the Galaxy started with the WMAP+SBBN lithium abundance, thus the evolution curve of lithium abundance stays well above the \textit{Spite Plateau} \citep{Matteucci,prantzos}. It has already been pointed out by \citet{francois} that the lithium abundance predicted by the model of \citet{prantzos} is about two times higher than that observed in NGC~2243. We may add here one further observation that while 47~Tuc lies at a metallicity at which, according to the model, Galactic cosmic rays are already contributing as much as 20\% of the primordial lithium abundance, it lies well below the other globular clusters. It should, however, be borne in mind that although 47~Tuc is as metal-rich as many disc stars, it is much older, therefore it could not have benefited of the lithium produced by cosmic ray spallation. Similarly, it may be simply because of its rather old age that the lithium abundance in NGC~2243 is significantly lower than that predicted by the models of Galactic chemical evolution.

\begin{figure}
\centering
\includegraphics[width=\columnwidth]{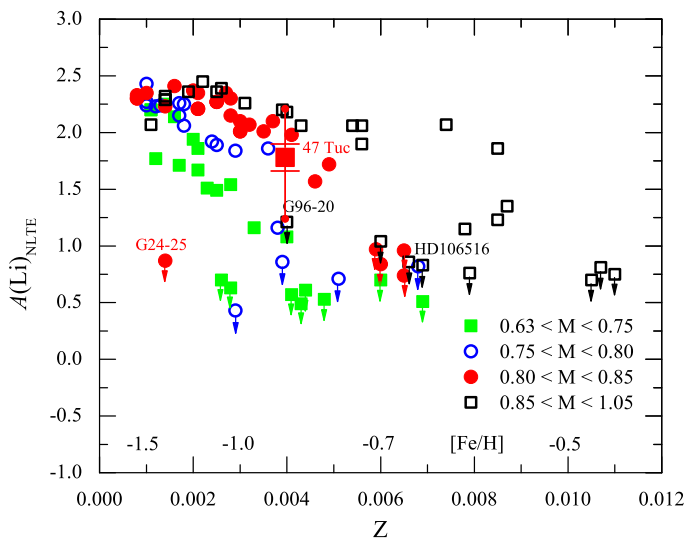}
 \caption{Lithium as a function of $Z$ for the sample of halo stars of \citet{NS12}, equivalent to Fig. 3 of \citet{NS12}, to which we added our lithium abundances for 47 Tuc. In the latter case, the symbol for 47~Tuc has the same meaning as in the case of the globular cluster symbols shown in Fig.~\ref{fig:plot_GC}.}
\label{fig:li_z}
\end{figure}

\subsection{Comparison with the halo field stars}

We also compare our results for 47 Tuc with those obtained in the recent investigation of \citet{NS12} who measured lithium abundances in a sample of halo stars in the metallicity range $-1.4<$[Fe/H]$<-0.7$. In terms of \FeH, 47 Tuc is therefore just on the edge of this sample, although it is well within the $Z$ range (see below). The first important result of \citet{NS12} is that there is no discernible difference in the lithium abundances of stars with different content of $\alpha$ elements. The second important result is that they highlight an almost linear dependence of Li abundance on both stellar mass and metallicity, with the less massive stars and the more metal-rich ones having lower Li abundances. Qualitatively, this result is indeed expected, because of the deepening of the convective envelope for less massive and more metal rich stars that is predicted by models of stellar evolution. On the $Z$ scale adopted by \citet{NS12}, based on Table 2 of \citet{y2}, 47 Tuc has $Z=0.00396$.

In Fig.\,\ref{fig:li_z} we reproduce Fig.~3 of \citet{NS12}, adding to it our lithium measurements for 47 Tuc. We may see that only a fraction of stars in the cluster have a lithium abundance comparable to that of field stars of the same mass and $Z$, while many others show lower lithium abundances. We believe the most obvious interpretation of this fact is that the most Li-rich stars in the cluster have suffered the same Li-depletion as field stars of the same mass and $Z$, while the others have suffered additional Li-depletion. This could be understood if the Na-O anticorrelation  and the possible Li-Na anticorrealtion are interpreted in terms of multiple stellar populations and interstellar medium pollution by the first generations of stars.

It may be interesting to note that the linear relations of \citet{NS12} cannot be extrapolated to lower metallicities, since they would predict higher Li abundances than those observed in field stars, as is nicely demonstrated by their Fig.~7. It  thus seems  that these relations describe accurately lithium depletion over the metallicity range for which they have been determined. It would be therefore interesting to compare them with the data of globular clusters M\,4 and NGC\,6752, that lie at the low metallicity edge of the sample studied by \citet{NS12}.

\subsection{Possible role of diffusion}

The scenario we have described in Sect.~\ref{clusters} assumes that the initial lithium abundance in Galactic globular clusters and metal-poor field stars was close to the value of the \textit{Spite Plateau}. While it could be also possible to conceive that it had the initial WMAP+SBBN value instead, a very contrived mechanism for the Li-depletion must be sought for in order to explain, simultaneously: \textit{(i)} the \textit{Spite Plateau}, \textit{(ii)} the lower abundance of 47~Tuc, and \textit{(iii)} the depletion of lithium, with respect to the predictions of chemical evolution models, in NGC~2243.

\citet{Korn06,Korn07}, from a careful study of stars in NGC\,6397, found significant differences in the Fe abundance between the TO stars and
the stars at the base of the RGB, in the sense that the iron abundance is lower at the TO and higher at the base of the RGB. They interpreted this as evidence of atomic diffusion that allows iron to settle  below the observable zone during the Main Sequence phase, but then to be brought back again to the atmosphere as the star expands and cools while evolving along the sub-giant branch (SGB). The models for turbulent diffusion of \citet{Richard} seem capable to reproduce the observed trends. \citet{Korn06} suggested that this process is indeed the one that lowers the lithium from the  WMAP+SBBN value to what is currently observed in globular clusters and metal-poor field stars. These models predict also an increase in the Li abundance along the SGB until dilution becomes important and the Li abundance rapidly decreases. This behaviour is indeed observed by
\citet{Korn06,Korn07} and confirmed in subsequent studies by the same group \citep{Lind09,Nordlander}. The differences in abundances for other elements are also supported by the analysis of \citet{Lind08}.

It should be noted however that other groups obtain different results for this same cluster. \citet{KochMc} do not find any significant difference in iron abundance between TO and RGB stars, although \citet{Nordlander} point out that a difference is visible, as long as one considers only the \ion{Fe}{ii} lines. It should be born in mind that the \ion{Fe}{ii} lines are sensitive to gravity and any error in the determination of surface gravities for the giant stars will affect the derived abundance. \citet{GHB09} analyzed a sample of MS and SGB stars in NGC\,6397 spanning the same range in colours, therefore effective temperatures, in order to minimise the uncertainties in comparing the Li abundances. They were able to demonstrate that the Li abundances are lower in MS stars than in SGB stars. However, both sets of stars, in their analysis displayed a steady decrease in lithium abundances with decreasing effective temperatures. This is at odds with the results of \citet{Lind09}. The fact that all the spectra of \citet{GHB09} have also been analyzed by \citet{Lind09} shows how delicate this analysis is. The differences in Li abundances between the two studies is essentially rooted in the different effective temperatures adopted.

In the globular cluster M\,4, \citet{Mucciarelli} measured lithium abundances in the TO stars and those along the SGB, and found that the lithium abundance steadily decreases as temperature decreases, without ever observing an increase in lithium. This behaviour is qualitatively identical to what was found by \citet{GHB09} in NGC\,6397, but at odds with the findings of \citet{Lind09} for the same cluster.

Collectively, the above-mentioned studies, in spite of some contradictions among them, seem to suggest strongly that some mechanism may alter slightly the atmospheric chemical abundances along the evolution of a star. This is not at odds with the scenario we proposed in Sect.\,\ref{clusters}. Whether such processes are capable of diminishing the lithium abundances from the primordial WMAP+SBBN value to what is currently observed is, in our view, more doubtful. The only models that currently claim to be able to do so are the turbulent diffusive models of \citet{Richard}. These models are parametrised, and the parameter governing the efficiency of diffusion is not based on any physical consideration, but is adjusted to fit the observations. It is somewhat worrisome that to explain their observations in NGC\,6752 with these models,  \citet{Gruyters} needed to invoke a different parameter, and less efficient mixing than what is invoked in NGC\,6397. In Fig.\,\ref{fig:plot_GC} the, already noted, near constancy of the mean Li abundance in the different low metallicity clusters, would then require the existence of some extremely fine-tuned mechanism that would imply different mixing efficiencies in the clusters of different metallicities in such a way as to allow them to deplete Li by nearly exactly the same amount, at the same time acting differently on other elements, like iron or magnesium. At higher metallicities, the same mechanism should be again tuned in order to match our observations of 47 Tuc. This is, indeed, possible, but it seems rather contrived. Another strong argument against diffusion as a viable and only mechanism for depleting lithium from the WMAP+SBBN value to the observed value is the near-constancy of the lithium abundances in $\omega$ Cen found by \citet{Monaco10} for stars that span an age range of 5.6 Gyr. Since diffusion is necessarily a time-dependent phenomenon, one would expect to observe different degrees of depletion, the largest ones for the oldest stars, but this is not, apparently, the case.

Obviously, additional observational studies of these effects are highly desirable, with special care taken to keep all systematic effects under control. Nevertheless, we believe a major breakthrough may only come together with a better theoretical understanding of the physical processes involved. Current simple parametric models have little predictive power, and more sophisticated and physically motived models are needed.

\section{Conclusions\label{sect:conclusions}}

We determined abundance of lithium in the atmospheres of 94 TO stars, as well as those of oxygen, and sodium in 110 TO stars belonging to the globular cluster 47~Tuc, taking into account NLTE and 3D hydrodynamical effects. The departures from LTE play a dominant role in the abundance determination: differences in the derived abundances reach to $\Delta_{\rm 1D\,NLTE-LTE} \approx -0.35$\,dex in the case of sodium and to $\Delta_{\rm 1D\,NLTE-LTE} \approx -0.20$\,dex in the case of oxygen. The role of convection in the atmospheres of TO stars in 47~Tuc plays a much lesser role in the spectral line formation, which leads to significantly smaller abundance corrections of $\Delta_{\rm 3D\,LTE - 1D\,LTE} \approx +0.02$\,dex for sodium and $\Delta_{\rm 3D\,LTE - 1D\,LTE} \approx +0.05$\,dex for oxygen.

Sodium and oxygen abundances are anti-correlated and our result is in very good agreement with that obtained by \citet{DLG10}. On the other hand, the average lithium abundance obtained in our study, $\langle A({\rm Li})\rangle = 1.78 \pm 0.18$, is $\approx0.27$~dex lower than that determined by \citet{DLG10}. Our data also hints towards a possible existence of Li--O correlation.

The mean Li abundance in 47~Tuc is lower than what is observed in other Galactic globular clusters by roughly a factor of three. The highest Li abundance observed in 47~Tuc is {\em lower} than the lowest Li abundances observed in the open cluster NGC~2243, that is only 0.2~dex more metal rich but 8~Gyr younger than 47~Tuc. When put into context with Li observations in other globular and open clusters and field stars, our results suggest a scenario in which the Li depletion during the star's MS/TO/SGB lifetime, is essentially zero for stars of metallicity lower than about $-1.0$, and becomes more important as soon metallicity increases. The initial lithium abundance with which the stars were created may in fact depend only on their age (it is larger for the younger stars) and not on their metallicity. These facts may explain in a natural way, for example, the difference in the lithium content between 47 Tuc and NGC\,2243.

To confirm (or disprove) the proposed scenario it would be important to observe lithium in other Galactic globular clusters of metallicity around $-1.0$, which may also allow determination of the metallicity at which the convective envelope becomes deep enough to result in significant lithium depletion.

\begin{acknowledgements}
We thank Valentina D'Orazi for providing details of her own analysis and useful discussions during our analysis of the 47~Tuc data. We thank the referee Andreas Korn for his useful comments which helped to improve the paper significantly. This work was supported by grants from the Research Council of Lithuania (MIP-065/2013 and TAP LZ 06/2013). EC, HGL, AK, and LS acknowledge financial support by the Sonderforschungsbereich SFB\,881 ``The Milky Way System'' (subproject A4 and A5) of the German Research Foundation (DFG). HGL, SAK, and MS acknowledge funding from the Research Council of Lithuania for the research visits to Vilnius. The study was based on observations made with the European Southern Observatory telescopes obtained from the ESO/ST-ECF Science Archive Facility.
\end{acknowledgements}

\bibliographystyle{aa}


\begin{appendix}

\section{Atmospheric parameters of TO stars in 47~Tuc and derived abundances of Li, O, and Na\label{sect:AppA}}

In this Section we provide the determined atmospheric parameters of the studied turn-off (TO) stars in 47~Tuc, as well as the abundances of lithium, oxygen, and sodium derived in their atmospheres. The abundances were derived assuming identical microturbulence velocity of $\xi_{\rm micro}=1.0 {\rm km\,s}^{-1}$ for all stars. The contents of the Table~\ref{tab:Tuc47-atmpar} are as follows: Column~1: star ID; Columns~2 and 3: right accension and declination; Columns~4 and 5: $B$ and $V$ magnitudes; Columns 6~and 7: effective temperature and surface gravity; Column~8: 3D~NLTE abundance of Li; Columns~9 and 11: 1D~NLTE abundances of O and Na, respectively; Columns~10 and 12: 3D+NLTE abundances of O and Na, respectively (i.e., 1D~NLTE abundances with 3D--1D abundance corrections taken into account).

\onecolumn
\begin{longtab}
\setlength{\tabcolsep}{3pt}
\begin{longtable}{lccccccccccc}
\caption{List of the investigated TO stars in 47~Tuc, their adopted atmospheric parameters and determined abundances of Li, O, and Na. \label{tab:Tuc47-atmpar} } \\
\hline\hline
ID & RA(2000)  & Dec(2000)   & $B$  & $V$    & \logg  & \teff & \textit{A}(Li) & \textit{A}(O) & \textit{A}(O) & \textit{A}(Na) & \textit{A}(Na) \\
            & deg           & deg             & mag   & mag     &  [cgs]   & K   & 3D~NLTE & 1D~NLTE & 3D+NLTE & 1D~NLTE  & 3D+NLTE  \\
\hline
\endfirsthead
\caption{continued.}\\
\hline
ID & RA(2000)  & Dec(2000)   & $B$  & $V$    & \logg  & \teff & \textit{A}(Li) & \textit{A}(O) & \textit{A}(O) & \textit{A}(Na) & \textit{A}(Na)\\
            & deg           & deg             & mag   & mag     &  [cgs]   & K   & 3D NLTE & 1D NLTE & 3D+NLTE & 1D NLTE  & 3D+NLTE  \\
\hline
\endhead
\hline
\endfoot
00006129  & 6.15846 & $-71.96322$ & 17.932 & 17.376 & 4.06 & 5851  & 1.61 & 8.36 & 8.42 & 5.78 & 5.81 \\
00006340  & 5.96746 & $-71.96075$ & 17.865 & 17.304 & 4.02 & 5817  & \ldots & 8.09 & 8.13 & 5.69 & 5.71 \\
00007619  & 6.33533 & $-71.94289$ & 17.953 & 17.392 & 4.05 & 5790  & 1.24 & 8.24 & 8.30 & 5.68 & 5.70 \\
00007969  & 6.11763 & $-71.93814$ & 17.977 & 17.409 & 4.06 & 5811  & 1.82 & 8.50 & 8.57 & 5.37 & 5.38 \\
00008359  & 6.24488 & $-71.93133$ & 17.941 & 17.369 & 4.06 & 5839  & 1.99 & 8.40 & 8.46 & 5.55 & 5.56 \\
00008881  & 6.16508 & $-71.92217$ & 17.970 & 17.409 & 4.10 & 5916  & 2.09 & 8.30 & 8.35 & 5.66 & 5.68 \\
00009191  & 6.21133 & $-71.91592$ & 17.965 & 17.400 & 4.07 & 5826  & 1.92 & 8.34 & 8.41 & 5.46 & 5.49 \\
00009243  & 6.27892 & $-71.91464$ & 17.978 & 17.421 & 4.08 & 5857  & 1.96 & 8.39 & 8.46 & 5.36 & 5.35 \\
00009434  & 6.24204 & $-71.91056$ & 17.959 & 17.399 & 4.08 & 5872  & 1.86 & 8.15 & 8.19 & 5.54 & 5.54 \\
00009540  & 6.11050 & $-71.90853$ & 17.933 & 17.366 & 4.06 & 5843  & 1.93 & 8.28 & 8.34 & 5.58 & 5.59 \\
00014912  & 5.80258 & $-71.96294$ & 17.949 & 17.397 & 4.08 & 5859  & 1.56 & 8.20 & 8.26 & 5.63 & 5.65 \\
00015086  & 5.76054 & $-71.96000$ & 17.950 & 17.382 & 4.07 & 5878  & 2.03 & 8.39 & 8.46 & 5.43 & 5.44 \\
00015174  & 5.82779 & $-71.95847$ & 17.936 & 17.370 & 4.04 & 5788  & 1.66 & 8.28 & 8.34 & 5.56 & 5.57 \\
00015346  & 5.58437 & $-71.95531$ & 17.968 & 17.382 & 4.03 & 5725  & 1.70 & 8.50 & 8.57 & 5.45 & 5.45 \\
00016131  & 5.77725 & $-71.94094$ & 18.000 & 17.426 & 4.07 & 5823  & 1.71 & 8.41 & 8.48 & 5.44 & 5.46 \\
00016631  & 5.75729 & $-71.92917$ & 18.010 & 17.441 & 4.08 & 5820  & 1.77 & 8.42 & 8.49 & 5.49 & 5.49 \\
00017628  & 5.87896 & $-71.90236$ & 17.997 & 17.432 & 4.11 & 5925  & 1.83 & 8.05 & 8.09 & 5.56 & 5.58 \\
00017767  & 5.84779 & $-71.89828$ & 17.953 & 17.385 & 4.06 & 5812  & 1.77 & 8.11 & 8.16 & 5.76 & 5.79 \\
00031830  & 5.41504 & $-72.04769$ & 17.925 & 17.357 & 4.05 & 5832  & 1.95 & 8.26 & 8.32 & 5.49 & 5.50 \\
00036086  & 5.70875 & $-72.20400$ & 17.918 & 17.350 & 4.05 & 5850  & 1.92 & 8.08 & 8.12 & 5.69 & 5.72 \\
00036747  & 5.77333 & $-72.19608$ & 18.007 & 17.463 & 4.09 & 5814  & \ldots & 8.06 & 8.11 & 5.61 & 5.62 \\
00038656  & 5.62004 & $-72.17497$ & 17.958 & 17.404 & 4.08 & 5850  & 1.78 & 8.20 & 8.25 & 5.55 & 5.56 \\
00040049  & 5.74092 & $-72.16181$ & 17.949 & 17.401 & 4.07 & 5822  & 1.78 & 8.01 & 8.05 & 5.64 & 5.66 \\
00040087  & 5.53888 & $-72.16119$ & 17.920 & 17.345 & 4.03 & 5787  & 1.80 & 8.51 & 8.59 & 5.51 & 5.51 \\
00040355  & 5.72492 & $-72.15906$ & 17.931 & 17.357 & 4.06 & 5879  & 1.78 & 8.29 & 8.34 & 5.47 & 5.49 \\
00043095  & 5.67775 & $-72.13700$ & 17.974 & 17.406 & 4.05 & 5770  & \ldots & 8.39 & 8.46 & 5.55 & 5.57 \\
00043108  & 5.57883 & $-72.13678$ & 17.894 & 17.330 & 4.03 & 5797  & \ldots & 8.19 & 8.25 & \ldots & \ldots \\
00044983  & 5.71950 & $-72.12375$ & 17.895 & 17.330 & 4.04 & 5848  & 1.88 & 8.34 & 8.41 & 5.53 & 5.53 \\
00045982  & 5.64500 & $-72.11706$ & 17.901 & 17.326 & 4.00 & 5707  & 1.85 & 8.38 & 8.44 & 5.50 & 5.51 \\
00046498  & 5.51050 & $-72.11339$ & 17.928 & 17.354 & 4.04 & 5790  & 1.82 & 8.47 & 8.54 & 5.33 & 5.33 \\
00049829  & 5.76571 & $-72.09175$ & 17.879 & 17.292 & 3.99 & 5740  & 1.65 & 8.43 & 8.49 & 5.34 & 5.35 \\
00051341  & 5.55921 & $-72.08197$ & 17.931 & 17.330 & 4.01 & 5731  & \ldots & 8.43 & 8.50 & 5.44 & 5.43 \\
00051740  & 5.53704 & $-72.07939$ & 17.953 & 17.383 & 4.07 & 5857  & 1.73 & 8.29 & 8.35 & 5.48 & 5.51 \\
00052108  & 5.50988 & $-72.07694$ & 17.909 & 17.328 & 3.99 & 5688  & 1.84 & 8.50 & 8.58 & 5.42 & 5.42 \\
00054596  & 5.61767 & $-72.06100$ & 17.932 & 17.370 & 4.05 & 5825  & 1.85 & 8.35 & 8.41 & 5.46 & 5.47 \\
00058492  & 5.68208 & $-72.03306$ & 17.947 & 17.366 & 4.02 & 5728  & 1.64 & 8.31 & 8.37 & 5.46 & 5.46 \\
00059579  & 5.66825 & $-72.02414$ & 17.895 & 17.324 & 3.98 & 5660  & 1.87 & 8.34 & 8.40 & 5.37 & 5.37 \\
00061639  & 5.69313 & $-72.00528$ & 17.891 & 17.340 & 4.03 & 5779  & 1.91 & 8.35 & 8.41 & 5.21 & 5.20 \\
00062314  & 5.56467 & $-71.99794$ & 18.006 & 17.384 & 4.03 & 5740  & \ldots & 8.42 & 8.49 & 5.43 & 5.44 \\
00062737  & 5.58004 & $-71.99319$ & 17.943 & 17.358 & 4.01 & 5691  & \ldots & 8.07 & 8.12 & 5.64 & 5.66 \\
00062773  & 5.87338 & $-71.99317$ & 17.883 & 17.331 & 4.05 & 5854  & 1.90 & 8.12 & 8.16 & 5.73 & 5.75 \\
00063201  & 5.60025 & $-71.98767$ & 17.927 & 17.335 & 4.02 & 5759  & 1.73 & 8.41 & 8.48 & 5.36 & 5.36 \\
00063954  & 5.77167 & $-71.97908$ & 17.902 & 17.309 & 4.02 & 5801  & 1.54 & 8.06 & 8.10 & 5.65 & 5.67 \\
00063973  & 5.70850 & $-71.97875$ & 17.899 & 17.314 & 4.02 & 5780  & 1.86 & 8.29 & 8.35 & 5.40 & 5.40 \\
00065981  & 6.05225 & $-72.22219$ & 17.972 & 17.424 & 4.07 & 5814  & \ldots & 8.21 & 8.27 & 5.69 & 5.71 \\
00066603  & 6.05375 & $-72.21225$ & 17.973 & 17.426 & 4.08 & 5848  & 1.77 & 8.03 & 8.06 & 5.68 & 5.70 \\
00066813  & 6.34237 & $-72.20903$ & 17.965 & 17.409 & 4.07 & 5817  & 1.61 & 8.27 & 8.32 & 5.44 & 5.45 \\
00066840  & 6.25950 & $-72.20878$ & 18.043 & 17.502 & 4.10 & 5780  & 1.84 & 8.15 & 8.20 & 5.60 & 5.61 \\
00067280  & 6.02708 & $-72.20253$ & 17.979 & 17.430 & 4.07 & 5808  & 1.77 & 8.24 & 8.29 & 5.44 & 5.45 \\
00069585  & 6.29904 & $-72.17517$ & 17.950 & 17.392 & 4.08 & 5888  & 2.00 & 8.32 & 8.38 & 5.38 & 5.36 \\
00070686  & 6.22921 & $-72.16494$ & 17.937 & 17.370 & 4.05 & 5808  & 1.64 & 8.13 & 8.17 & 5.45 & 5.45 \\
00070910  & 6.27663 & $-72.16297$ & 17.949 & 17.404 & 4.06 & 5797  & 1.78 & 8.22 & 8.27 & 5.54 & 5.55 \\
00071404  & 6.29454 & $-72.15886$ & 17.976 & 17.409 & 4.06 & 5787  & \ldots & 8.33 & 8.39 & 5.79 & 5.81 \\
00072011  & 6.11733 & $-72.15458$ & 18.021 & 17.446 & 4.05 & 5702  & 1.43 & 8.42 & 8.49 & 5.51 & 5.52 \\
00096225  & 6.27933 & $-72.02936$ & 17.898 & 17.339 & 4.04 & 5805  & 1.86 & 8.40 & 8.47 & 5.35 & 5.35 \\
00097156  & 6.36075 & $-72.02406$ & 17.930 & 17.360 & 4.03 & 5750  & 1.70 & 8.43 & 8.50 & 5.48 & 5.49 \\
00099636  & 6.26008 & $-72.00881$ & 17.942 & 17.404 & 4.06 & 5799  & 1.48 & 8.25 & 8.31 & 5.57 & 5.58 \\
00100325  & 6.35675 & $-72.00369$ & 17.939 & 17.377 & 4.05 & 5794  & 1.70 & 8.39 & 8.45 & 5.52 & 5.53 \\
00102294  & 6.06792 & $-71.98808$ & 17.887 & 17.315 & 4.01 & 5772  & 1.92 & 8.40 & 8.47 & 5.36 & 5.37 \\
00102307  & 6.21471 & $-71.98781$ & 17.937 & 17.388 & 4.07 & 5835  & 1.64 & 8.34 & 8.41 & 5.40 & 5.40 \\
00103067  & 5.94763 & $-71.98056$ & 17.878 & 17.314 & 3.98 & 5665  & \ldots & 8.16 & 8.21 & 5.71 & 5.73 \\
00103709  & 6.02521 & $-71.97353$ & 17.893 & 17.312 & 4.02 & 5806  & 1.66 & 8.14 & 8.18 & 5.64 & 5.66 \\
00104049  & 6.17200 & $-71.96964$ & 17.894 & 17.324 & 4.02 & 5768  & 1.67 & 8.10 & 8.14 & 5.67 & 5.69 \\
00106794  & 6.47321 & $-72.18328$ & 18.016 & 17.459 & 4.08 & 5789  & 1.74 & 8.34 & 8.40 & 5.53 & 5.52 \\
00107260  & 6.45896 & $-72.17064$ & 17.922 & 17.377 & 4.06 & 5829  & 1.73 & 8.34 & 8.41 & 5.33 & 5.33 \\
00107528  & 6.57650 & $-72.16361$ & 17.983 & 17.441 & 4.11 & 5923  & 2.02 & 8.38 & 8.45 & 5.55 & 5.55 \\
00107618  & 6.59092 & $-72.16119$ & 17.923 & 17.370 & 4.06 & 5831  & 2.03 & 8.36 & 8.42 & 5.57 & 5.58 \\
00107866  & 6.56246 & $-72.15469$ & 18.006 & 17.461 & 4.06 & 5727  & 1.78 & 8.51 & 8.58 & 5.14 & 5.12 \\
00108171  & 6.40738 & $-72.14778$ & 17.938 & 17.388 & 4.06 & 5812  & \ldots & 7.92 & 7.95 & 5.84 & 5.86 \\
00108389  & 6.58104 & $-72.14253$ & 17.933 & 17.382 & 4.05 & 5808  & 1.61 & 8.22 & 8.27 & 5.51 & 5.53 \\
00109441  & 6.53275 & $-72.11814$ & 17.919 & 17.361 & 4.07 & 5875  & 1.70 & 8.22 & 8.27 & 5.42 & 5.41 \\
00109777  & 6.50933 & $-72.11058$ & 17.959 & 17.415 & 4.09 & 5873  & 2.00 & 8.36 & 8.42 & 5.54 & 5.55 \\
00110197  & 6.60008 & $-72.10139$ & 17.977 & 17.425 & 4.08 & 5824  & 1.78 & 8.60 & 8.68 & \ldots & \ldots \\
00111136  & 6.48775 & $-72.08114$ & 17.928 & 17.363 & 4.08 & 5908  & 1.98 & 8.36 & 8.42 & 5.40 & 5.42 \\
00111231  & 6.52613 & $-72.07919$ & 17.945 & 17.374 & 4.03 & 5732  & 1.66 & 8.41 & 8.47 & 5.28 & 5.28 \\
00112473  & 6.59492 & $-72.05506$ & 17.903 & 17.357 & 4.05 & 5840  & 1.67 & 8.34 & 8.40 & 5.45 & 5.44 \\
00112684  & 6.46096 & $-72.05136$ & 17.892 & 17.339 & 4.03 & 5780  & 1.55 & 8.30 & 8.36 & 5.39 & 5.39 \\
00113090  & 6.47025 & $-72.04353$ & 17.956 & 17.391 & 4.05 & 5794  & 1.64 & 8.32 & 8.39 & 5.47 & 5.48 \\
00113841  & 6.55175 & $-72.02800$ & 17.952 & 17.394 & 4.07 & 5854  & 1.96 & 8.52 & 8.60 & 5.44 & 5.44 \\
00113959  & 6.49396 & $-72.02594$ & 17.934 & 17.382 & 4.10 & 5968  & \ldots & 8.35 & 8.41 & 5.48 & 5.48 \\
00115880  & 6.51471 & $-71.98331$ & 17.940 & 17.372 & 4.06 & 5845  & 2.21 & 8.30 & 8.35 & 5.51 & 5.53 \\
10000002  & 5.43304 & $-72.05411$ & 17.963 & 17.406 & 4.09 & 5894  & 1.85 & 8.37 & 8.43 & 5.34 & 5.34 \\
10000004  & 5.62229 & $-72.10428$ & 17.946 & 17.398 & 4.10 & 5934  & 2.07 & 8.37 & 8.43 & 5.43 & 5.43 \\
10000008  & 5.70025 & $-72.15828$ & 18.059 & 17.499 & 4.13 & 5905  & 2.04 & 8.09 & 8.14 & 5.53 & 5.54 \\
10000009  & 5.70075 & $-72.09483$ & 18.038 & 17.482 & 4.09 & 5792  & 1.65 & 8.52 & 8.59 & 5.51 & 5.50 \\
10000012  & 5.70475 & $-72.08533$ & 17.855 & 17.298 & 4.03 & 5836  & 1.95 & 8.38 & 8.44 & 5.42 & 5.42 \\
10000015  & 5.72129 & $-72.07636$ & 17.912 & 17.339 & 4.02 & 5754  & 1.51 & 8.48 & 8.55 & 5.46 & 5.46 \\
10000016  & 5.72533 & $-72.02817$ & 17.879 & 17.283 & 3.98 & 5724  & 1.90 & 8.36 & 8.43 & 5.31 & 5.31 \\
10000020  & 5.75263 & $-72.06483$ & 18.002 & 17.464 & 4.10 & 5834  & 1.65 & 8.11 & 8.15 & 5.59 & 5.61 \\
10000022  & 5.76167 & $-72.04869$ & 17.851 & 17.276 & 3.99 & 5749  & 1.40 & 8.16 & 8.22 & 5.68 & 5.70 \\
10000026  & 5.77117 & $-72.12517$ & 17.972 & 17.411 & 4.06 & 5784  & 1.76 & 8.35 & 8.41 & 5.54 & 5.55 \\
10000027  & 5.77721 & $-72.12919$ & 18.003 & 17.444 & 4.08 & 5829  & 1.73 & 8.29 & 8.35 & 5.45 & 5.46 \\
10000036  & 5.84604 & $-72.00550$ & 17.902 & 17.337 & 4.00 & 5706  & 1.63 & 7.84 & 7.87 & 5.65 & 5.66 \\
10000038  & 5.86846 & $-72.19789$ & 17.838 & 17.254 & 4.00 & 5810  & 1.82 & 8.34 & 8.40 & 5.53 & 5.53 \\
10000041  & 5.90950 & $-71.93806$ & 17.933 & 17.378 & 4.08 & 5889  & 2.11 & 8.44 & 8.51 & 5.46 & 5.47 \\
10000043  & 5.94513 & $-72.17733$ & 17.955 & 17.372 & 4.05 & 5883  & 1.73 & 8.18 & 8.23 & 5.85 & 5.88 \\
10000048  & 5.99058 & $-71.98381$ & 17.930 & 17.368 & 4.05 & 5832  & 1.76 & 8.14 & 8.18 & 5.64 & 5.65 \\
10000049  & 6.00479 & $-72.18656$ & 18.036 & 17.498 & 4.14 & 5935  & 2.01 & 8.32 & 8.38 & 5.54 & 5.55 \\
10000053  & 6.04242 & $-72.20942$ & 17.936 & 17.384 & 4.08 & 5881  & \ldots & 8.08 & 8.12 & 5.68 & 5.70 \\
10000057  & 6.08746 & $-71.93789$ & 18.028 & 17.478 & 4.10 & 5846  & 1.76 & 8.21 & 8.26 & 5.70 & 5.72 \\
10000062  & 6.12154 & $-71.97469$ & 17.987 & 17.434 & 4.10 & 5891  & \ldots & 7.73 & 7.75 & 5.76 & 5.79 \\
10000068  & 6.15775 & $-71.95836$ & 17.956 & 17.388 & 4.09 & 5923  & \ldots & 8.03 & 8.07 & 5.74 & 5.78 \\
10000072  & 6.19088 & $-71.97972$ & 18.012 & 17.464 & 4.09 & 5829  & \ldots & 8.12 & 8.16 & 5.60 & 5.61 \\
10000073  & 6.21196 & $-72.00553$ & 17.854 & 17.293 & 4.03 & 5855  & 1.33 & 8.27 & 8.32 & 5.31 & 5.32 \\
10000075  & 6.24354 & $-71.96136$ & 17.951 & 17.388 & 4.08 & 5882  & 1.88 & 8.43 & 8.49 & 5.46 & 5.46 \\
10000079  & 6.27275 & $-72.12033$ & 18.036 & 17.501 & 4.09 & 5750  & 1.68 & 8.52 & 8.59 & 5.31 & 5.31 \\
10000086  & 6.30192 & $-72.05958$ & 17.868 & 17.298 & 3.99 & 5708  & 1.57 & 8.11 & 8.16 & 5.35 & 5.34 \\
10000088  & 6.31217 & $-72.03944$ & 18.008 & 17.458 & 4.07 & 5771  & 1.50 & 7.89 & 7.92 & 5.78 & 5.80 \\
10000090  & 6.34033 & $-71.96881$ & 17.928 & 17.363 & 4.08 & 5921  & 2.02 & 8.31 & 8.37 & 5.43 & 5.44 \\
10000094  & 6.42554 & $-72.07425$ & 18.013 & 17.460 & 4.10 & 5869  & 1.63 & 8.34 & 8.40 & 5.35 & 5.35 \\
 \hline
\end{longtable}
\end{longtab}
\twocolumn

\section{On the relationship between 3D~NLTE, 1D~NLTE, and 3D~LTE abundance corrections for the Li 670.8~nm line \label{sect:Li-test}}

\begin{table*}[tb]
 \begin{center}
\caption{Abundance corrections for the lithium doublet at 670.8~nm computed using model atmosphere with \teff = 5930~K, \logg = 4.0, \MoH = 0.0.
\label{tab:Li_dabu_t59g40mm00}}
 \vspace{4mm}
  \begin{tabular}{lccc}
  \hline\hline
\noalign{\smallskip}
 \textit{EW},   & \multicolumn{3}{c}{$\Delta$, dex} \\
  pm            & ${\rm 3D}^{\rm NLTE} - {\rm 1D}^{\rm LTE}$ & ${\rm 1D}^{\rm NLTE} - {\rm 1D}^{\rm LTE}$ & ${\rm 3D}^{\rm LTE} - {\rm 1D}^{\rm LTE}$ \\
 \hline\noalign{\smallskip}
 0.5  & 0.091  &   0.079  & $-0.014$   \\
 9.0  & 0.105  &   0.084  & $-0.010$   \\
  \hline
  \end{tabular}
  \end{center}
\end{table*}

\begin{table*}[tb]
 \begin{center}
\caption{Abundance corrections for the lithium doublet at 670.8~nm computed using model atmosphere with \teff = 5850~K, \logg = 4.0, \MoH = --1.0.
\label{tab:Li_dabu_t59g40mm10}}
 \vspace{4mm}
  \begin{tabular}{lccc}
  \hline\hline
\noalign{\smallskip}
 \textit{EW},   & \multicolumn{3}{c}{$\Delta$, dex} \\
  pm            & ${\rm 3D}^{\rm NLTE} - {\rm 1D}^{\rm LTE}$ &  ${\rm 1D}^{\rm NLTE} - {\rm 1D}^{\rm LTE}$ & ${\rm 3D}^{\rm LTE} - {\rm 1D}^{\rm LTE}$ \\
 \hline\noalign{\smallskip}
 0.5  & 0.065  &   0.041  & $-0.135$   \\
 9.0  & 0.065  &   0.037  & $-0.137$   \\
  \hline
  \end{tabular}
  \end{center}
\end{table*}

In the derivation of oxygen and sodium abundances we corrected the 1D~NLTE abundances for 3D hydrodynamical effects by adding $\Delta_{\rm 3D\,LTE - 1D\,LTE}$ abundance correction. It is obvious that such procedure does not account for the strong dependence of atomic level departure coefficients on the horizontal fluctuations of temperature in the 3D hydrodynamical models. Ideally, elemental abundances should be derived using full 3D~NLTE approach where NLTE spectral synthesis computations are performed in the framework of 3D hydrodynamical models. This, however, is still rarely possible since the majority of current 3D spectral synthesis codes lack the capability of NLTE radiative transfer computations. One is then, therefore, forced to resort to different simplifications, e.g., such as applying $\Delta_{\rm 1D\,NLTE - 1D\,LTE}+\Delta_{\rm 3D\,LTE - 1D\,LTE}$ abundance correction to take into account \textit{both} 3D hydrodynamical and NLTE effects.

It would be therefore instructive to assess whether the $\Delta_{\rm 3D\,NLTE - 1D\,LTE}$ and $\Delta_{\rm 1D\,NLTE - 1D\,LTE}+\Delta_{\rm 3D\,LTE - 1D\,LTE}$ corrections would lead to different final elemental abundances, and, if so, how large these differences may be expected to be. For this purpose, we used 3D hydrodynamical \COBOLD\ and 1D hydrostatic \LHD\ model atmospheres, together with the NLTE3D code, to compute full 3D, average $\xtmean{\mbox{3D}}$, and 1D synthetic profiles of the 670.8~nm resonance lithium doublet\footnote{NLTE3D code allows computation of atomic level population numbers for the model atom of Li in 3D~NLTE/LTE and 1D~NLTE/LTE \citep[see][for more details on the NLTE3D code]{SBC10,SCC12,PSK13}}. Spectral line synthesis was done with the \LINFOR\ code (see Sect.~\ref{sect:3Dabund}). Two 3D hydrodynamical models were used in this exercise (of the four utilised in Sect.~\ref{sect:3Dabund}), with the following atmospheric parameters: \teff = 5930~K, \logg = 4.0, $\MoH=0.0$, and \teff = 5850~K, \logg = 4.0, $\MoH=-1.0$. The analysis was done for the cases of (i) weak ($EW=0.5$\,pm), and (ii) strong ($EW=7-9$\,pm) lithium lines. The obtained synthetic line profiles were used to compute $\Delta_{\rm 3D\,NLTE - 1D\,LTE}$, $\Delta_{\rm 1D\,NLTE - 1D\,LTE}$, and $\Delta_{\rm 3D\,LTE - 1D\,LTE}$ abundance corrections. The resulting abundance corrections are provided in Tables~\ref{tab:Li_dabu_t59g40mm00} and \ref{tab:Li_dabu_t59g40mm10}.

The obtained results clearly show that full $\Delta_{\rm 3D\,NLTE - 1D\,LTE}$ correction is always different from the sum $\Delta_{\rm 1D\,NLTE - 1D\,LTE}+\Delta_{\rm 3D\,LTE - 1D\,LTE}$, both at $\MoH=0.0$ and $-1.0$. It is nevertheless important to stress that while at solar metallicity the abundances obtained with both approaches agree to $\approx0.03$~dex, significant differences are seen at $\MoH=-1.0$. In the latter case, the full 3D~NLTE correction is positive and amounts to $\Delta_{\rm 3D\,NLTE - 1D\,LTE}=0.065$\,dex for both weak and strong lines. The combined abundance correction, on the other hand, is negative and reaches to $\Delta_{\rm 1D\,NLTE - 1D\,LTE}+\Delta_{\rm 3D\,LTE - 1D\,LTE}=-0.094$~dex and $-0.100$~dex for weak and strong lines, respectively. Obviously, application of such combined correction in the case of lithium with the $\MoH=-1.0$ models may lead to erroneous results.

The reason why the differences between the two corrections become different at $\MoH=-1.0$ is that horizontal temperature fluctuations in the 3D hydrodynamical model \textit{and} differences between the temperature profiles of the average $\xtmean{\mbox{3D}}$ and 1D models become larger at lower metallicities. Since atomic population numbers depend very sensitively on the local temperature, this leads to stronger deviations from NLTE in the 3D models compared to what would be expected in the 1D case. In this sense, situation may be somewhat safer in the case of oxygen, the lines of which form deeper in the atmosphere and thus experience smaller temperature variations.

\begin{figure}[tb]
\centering
\includegraphics[width=8.85cm]{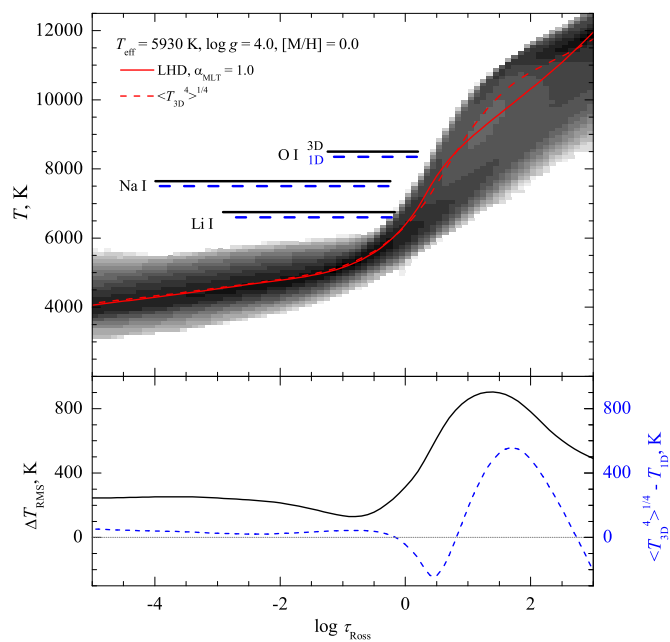}
 \caption{\textbf{Top:} temperature stratifications in the 3D hydrodynamical (grey scales indicating the temperature probability density), average $\xtmean{\mbox{3D}}$ (dashed red line), and 1D (solid red line) model atmospheres at $\MoH=0.0$. Horizontal bars show the optical depth intervals where 90\% of the line equivalent width is formed (i.e., 5\% to 95\%): black and dashed blue bars correspond to the line forming regions in the full 3D and 1D model atmospheres, respectively (the equivalent widths of Li, O, and Na lines are 2~pm, 4~pm, and 19~pm, respectively). \textbf{Bottom:} RMS value of horizontal temperature fluctuations in the 3D model (black line) and temperature difference between the average $\xtmean{\mbox{3D}}$ and 1D models (dashed blue line).}
 \label{fig:temp-stratm00}
\end{figure}

\begin{figure}[tb]
\centering
\includegraphics[width=8.85cm]{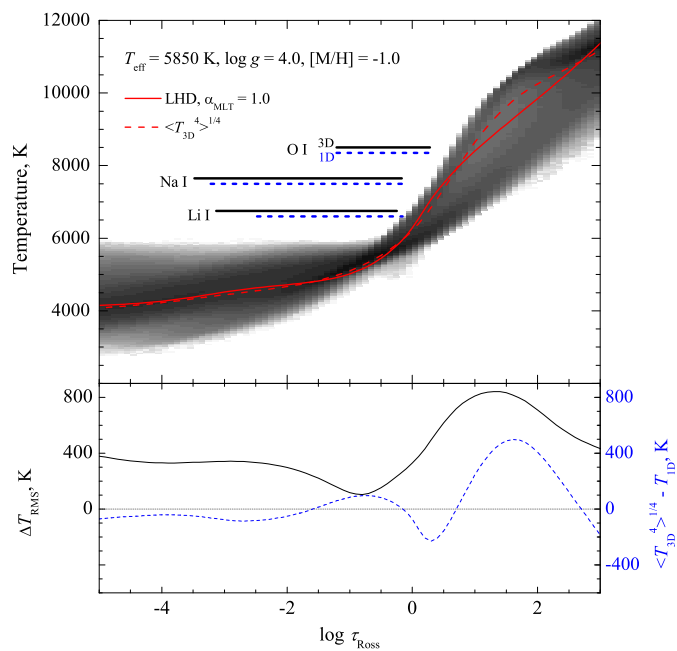}
 \caption{Same as in Fig.~\ref{fig:temp-stratm00} but for models at $\MoH=-1.0$.}
 \label{fig:temp-stratm10}
\end{figure}

\section{3D--1D LTE abundance corrections for oxygen and sodium}

In this Section we provide $\Delta_{\rm 3D\,LTE - 1D\,LTE}$ abundance corrections for the oxygen and sodium spectral lines used in this study. Abundance corrections are provided for all four models used in Sect.~\ref{sect:3Dabund}, and a range of line equivalent widths, $EW$.

The 3D-1D LTE abundance corrections provided in Tables~\ref{tab:dabut55g40mm00}--~\ref{tab:dabut59g40mm10} for $\xi_{\rm micro}=1.0{ \rm  km\,s}^{-1}$ were interpolated to the observed equivalent widths and added to the 1D~NLTE abundances to obtain the 3D+NLTE abundances for oxygen and sodium as given in Table~\ref{tab:Tuc47-atmpar}. As demonstrated in Appendix~\ref{sect:Li-test}, the 3D+NLTE abundances are not a good approximation to the real 3D~NLTE abundances in the case of Li (the 3D--1D LTE corrections do not even yield the correct sign). For oxygen and sodium, the validity of this approximation is unclear, and the corrections given in  Tables~\ref{tab:dabut55g40mm00}--~\ref{tab:dabut59g40mm10} should be considered as an order of magnitude estimate at best.

The corrections given for  $\xi_{\rm micro}=0.5$ and 1.5 ${\rm km\,s}^{-1}$ serve to demonstrate that the dependence of the 3D--1D LTE abundance corrections on the adopted microturbulence parameter is weak (see also Table~\ref{tab:abnsens}), even for stronger, partly saturated lines. This is explained by the fact that the thermal line broadening largely dominates over the turbulent broadening for these relatively light atoms. For heavier elements, like e.g. iron, the impact of microturbulence on the 3D abundance corrections would be significant, and a proper evaluation of the 3D corrections for stronger lines would require a more elaborate treatment than the simple assumption of a fixed microturbulence that was sufficient in the present work.

\begin{sidewaystable*}
\begin{center}
\caption{3D--1D abundance corrections of the spectral lines used in this work as a function of the equivalent width, {\it EW}, computed using the model atmosphere with $T_{\rm eff} = 5475$~K, $\log g = 4.0$,  $\MoH = 0.0$, and microturbulence velocities $\xi_{\rm micro} = 0.5, 1.0$, and 1.5  ${\rm km\, s^{-1} }$.
\label{tab:dabut55g40mm00}}
  \begin{tabular}{rccccccccccccccc}
  \hline\hline
\noalign{\smallskip}
{\it EW}, & \multicolumn{3}{c}{\ion{O}{i}} & \multicolumn{3}{c}{\ion{O}{i}}  & \multicolumn{3}{c}{\ion{O}{i}} & \multicolumn{3}{c}{\ion{Na}{i}} &  \multicolumn{3}{c}{\ion{Na}{i}} \\
 pm   & \multicolumn{3}{c}{777.2} &  \multicolumn{3}{c}{777.4} &  \multicolumn{3}{c}{777.5} & \multicolumn{3}{c}{818.3} &  \multicolumn{3}{c}{819.5} \\
      &  0.5  &  1.0  & 1.5 &  0.5  &  1.0  & 1.5  &  0.5  &  1.0  & 1.5 &  0.5  &  1.0  & 1.5 &  0.5  &  1.0  & 1.5 \\
\hline\noalign{\smallskip}
 0.4   & 0.001  & 0.004 & 0.007  & 0.001 & 0.003 & 0.007 & 0.001 & 0.003 & 0.007  &   0.008  & 0.009 & 0.011 &   0.008  &  0.009 & 0.011 \\
 0.6   & 0.005  & 0.009 & 0.015  & 0.005 & 0.009 & 0.014 & 0.006 & 0.009 & 0.014  &   0.008  & 0.009 & 0.012 &   0.008  &  0.009 & 0.011 \\
 0.8   & 0.009  & 0.014 & 0.021  & 0.009 & 0.014 & 0.021 & 0.009 & 0.014 & 0.021  &   0.007  & 0.009 & 0.012 &   0.007  &  0.009 & 0.012 \\
 1.0   & 0.013  & 0.018 & 0.027  & 0.013 & 0.019 & 0.027 & 0.012 & 0.018 & 0.027  &   0.006  & 0.009 & 0.013 &   0.006  &  0.009 & 0.013 \\
 1.2   & 0.016  & 0.023 & 0.032  & 0.016 & 0.023 & 0.033 & 0.015 & 0.022 & 0.032  &   0.006  & 0.009 & 0.014 &   0.006  &  0.009 & 0.014 \\
 1.5   & 0.019  & 0.028 & 0.040  & 0.019 & 0.027 & 0.039 & 0.019 & 0.027 & 0.039  &   0.005  & 0.009 & 0.015 &   0.004  &  0.009 & 0.015 \\
 2.0   & 0.023  & 0.034 & 0.049  & 0.023 & 0.034 & 0.049 & 0.023 & 0.034 & 0.049  &   0.003  & 0.009 & 0.018 &   0.003  &  0.009 & 0.017 \\
 2.5   & 0.027  & 0.039 & 0.057  & 0.027 & 0.039 & 0.057 & 0.026 & 0.038 & 0.056  &   0.000  & 0.009 & 0.020 &   0.001  &  0.009 & 0.020 \\
 3.0   & 0.029  & 0.043 & 0.063  & 0.029 & 0.043 & 0.063 & 0.029 & 0.043 & 0.062  & $-0.002$ & 0.009 & 0.022 & $-0.002$ &  0.010 & 0.024 \\
 3.5   & 0.031  & 0.046 & 0.068  & 0.031 & 0.046 & 0.068 & 0.031 & 0.046 & 0.068  & $-0.004$ & 0.009 & 0.026 & $-0.004$ &  0.010 & 0.026 \\
 4.0   & 0.033  & 0.049 & 0.072  & 0.033 & 0.049 & 0.072 & 0.033 & 0.049 & 0.072  & $-0.007$ & 0.010 & 0.030 & $-0.007$ &  0.009 & 0.029 \\
 4.5   & 0.034  & 0.051 & 0.075  & 0.034 & 0.051 & 0.075 & 0.035 & 0.051 & 0.076  & $-0.009$ & 0.011 & 0.035 & $-0.010$ &  0.009 & 0.033 \\
 5.0   & 0.035  & 0.053 & 0.078  & 0.036 & 0.053 & 0.079 & 0.036 & 0.053 & 0.079  & $-0.012$ & 0.011 & 0.039 & $-0.013$ &  0.010 & 0.037 \\
 5.5   & 0.037  & 0.055 & 0.081  & 0.037 & 0.055 & 0.081 & 0.037 & 0.055 & 0.081  & $-0.015$ & 0.011 & 0.042 & $-0.015$ &  0.011 & 0.042 \\
 6.0   & 0.039  & 0.057 & 0.083  & 0.039 & 0.057 & 0.083 & 0.038 & 0.056 & 0.082  & $-0.018$ & 0.011 & 0.047 & $-0.018$ &  0.012 & 0.048 \\
 6.5   & 0.040  & 0.058 & 0.085  & 0.040 & 0.058 & 0.085 & 0.040 & 0.058 & 0.084  & $-0.021$ & 0.012 & 0.052 & $-0.020$ &  0.013 & 0.055 \\
 7.0   & 0.041  & 0.059 & 0.087  & 0.041 & 0.059 & 0.086 & 0.041 & 0.059 & 0.086  & $-0.023$ & 0.013 & 0.057 & $-0.022$ &  0.014 & 0.060 \\
 7.5   & 0.042  & 0.060 & 0.088  & 0.042 & 0.060 & 0.087 & 0.042 & 0.060 & 0.088  & $-0.026$ & 0.014 & 0.063 & $-0.025$ &  0.015 & 0.065 \\
 8.0   & 0.043  & 0.061 & 0.088  & 0.043 & 0.061 & 0.088 & 0.044 & 0.062 & 0.089  & $-0.027$ & 0.016 & 0.069 & $-0.027$ &  0.016 & 0.070 \\
 8.5   & 0.044  & 0.062 & 0.089  & 0.044 & 0.062 & 0.089 & 0.045 & 0.062 & 0.089  & $-0.028$ & 0.018 & 0.076 & $-0.028$ &  0.017 & 0.075 \\
 9.0   & 0.045  & 0.062 & 0.089  & 0.046 & 0.064 & 0.091 & 0.045 & 0.063 & 0.090  & $-0.029$ & 0.020 & 0.082 & $-0.029$ &  0.018 & 0.080 \\
 9.5   & 0.046  & 0.063 & 0.090  & 0.047 & 0.064 & 0.091 & 0.046 & 0.064 & 0.090  & $-0.029$ & 0.022 & 0.088 & $-0.029$ &  0.020 & 0.085 \\
10.0   & 0.047  & 0.064 & 0.091  & 0.048 & 0.065 & 0.092 & 0.047 & 0.064 & 0.090  & $-0.029$ & 0.023 & 0.093 & $-0.029$ &  0.021 & 0.089 \\
11.0   & 0.049  & 0.066 & 0.092  & 0.049 & 0.066 & 0.092 & 0.049 & 0.065 & 0.091  & $-0.028$ & 0.026 & 0.100 & $-0.028$ &  0.026 & 0.098 \\
12.0   & 0.051  & 0.067 & 0.092  & 0.051 & 0.067 & 0.092 & 0.051 & 0.067 & 0.092  & $-0.024$ & 0.029 & 0.104 & $-0.024$ &  0.030 & 0.107 \\
13.0   & 0.052  & 0.068 & 0.092  & 0.052 & 0.067 & 0.092 & 0.052 & 0.068 & 0.092  & $-0.020$ & 0.032 & 0.106 & $-0.020$ &  0.034 & 0.111 \\
14.0   & 0.053  & 0.068 & 0.092  & 0.053 & 0.068 & 0.091 & 0.053 & 0.069 & 0.092  & $-0.016$ & 0.035 & 0.110 & $-0.015$ &  0.036 & 0.113 \\
15.0   & 0.054  & 0.069 & 0.091  & 0.054 & 0.069 & 0.091 & 0.055 & 0.069 & 0.092  & $-0.010$ & 0.038 & 0.112 & $-0.010$ &  0.037 & 0.111 \\
16.0   & \ldots & \ldots & \ldots & \ldots & \ldots & \ldots & \ldots & \ldots & \ldots & $-0.005$ & 0.041 & 0.112 & $-0.006$ &  0.039 & 0.109 \\
17.0   & \ldots & \ldots & \ldots & \ldots & \ldots & \ldots & \ldots & \ldots & \ldots &   0.000  & 0.043 & 0.111 & $-0.001$ &  0.042 & 0.108 \\
18.0   & \ldots & \ldots & \ldots & \ldots & \ldots & \ldots & \ldots & \ldots & \ldots &   0.004  & 0.044 & 0.108 &   0.003  &  0.043 & 0.107 \\
19.0   & \ldots & \ldots & \ldots & \ldots & \ldots & \ldots & \ldots & \ldots & \ldots &   0.007  & 0.045 & 0.105 &   0.007  &  0.045 & 0.106 \\
20.0   & \ldots & \ldots & \ldots & \ldots & \ldots & \ldots & \ldots & \ldots & \ldots &   0.011  & 0.046 & 0.103 &   0.011  &  0.046 & 0.104 \\
21.0   & \ldots & \ldots & \ldots & \ldots & \ldots & \ldots & \ldots & \ldots & \ldots &   0.014  & 0.047 & 0.101 &   0.014  &  0.047 & 0.101 \\
22.0   & \ldots & \ldots & \ldots & \ldots & \ldots & \ldots & \ldots & \ldots & \ldots &   0.017  & 0.048 & 0.099 &   0.017  &  0.048 & 0.099 \\
23.0   & \ldots & \ldots & \ldots & \ldots & \ldots & \ldots & \ldots & \ldots & \ldots &   0.019  & 0.049 & 0.097 &   0.019  &  0.048 & 0.096 \\
24.0   & \ldots & \ldots & \ldots & \ldots & \ldots & \ldots & \ldots & \ldots & \ldots &   0.021  & 0.049 & 0.095 &   0.022  &  0.049 & 0.093 \\
25.0   & \ldots & \ldots & \ldots & \ldots & \ldots & \ldots & \ldots & \ldots & \ldots &   0.024  & 0.050 & 0.093 &   0.023  &  0.049 & 0.091 \\
26.0   & \ldots & \ldots & \ldots & \ldots & \ldots & \ldots & \ldots & \ldots & \ldots &   0.025  & 0.050 & 0.091 &   0.025  &  0.049 & 0.089 \\
27.0   & \ldots & \ldots & \ldots & \ldots & \ldots & \ldots & \ldots & \ldots & \ldots &   0.027  & 0.051 & 0.089 &   0.027  &  0.049 & 0.087 \\
28.0   & \ldots & \ldots & \ldots & \ldots & \ldots & \ldots & \ldots & \ldots & \ldots &   0.029  & 0.051 & 0.087 &   0.028  &  0.050 & 0.085 \\
29.0   & \ldots & \ldots & \ldots & \ldots & \ldots & \ldots & \ldots & \ldots & \ldots &   0.030  & 0.050 & 0.085 &   0.029  &  0.050 & 0.084 \\
30.0   & \ldots & \ldots & \ldots & \ldots & \ldots & \ldots & \ldots & \ldots & \ldots &   0.030  & 0.050 & 0.082 &   0.030  &  0.050 & 0.082 \\
\hline
\end{tabular}
\end{center}
\end{sidewaystable*}

\begin{sidewaystable*}
\begin{center}
\caption{The same as in Table~\ref{tab:dabut55g40mm00} but for the model atmosphere with $T_{\rm eff} = 5533$~K, $\log g = 4.0$,  $\MoH = -1.0$.
\label{tab:dabut55g40mm10}}
  \begin{tabular}{rccccccccccccccc}
  \hline\hline
\noalign{\smallskip}
{\it EW}, & \multicolumn{3}{c}{\ion{O}{i}} & \multicolumn{3}{c}{\ion{O}{i}}  & \multicolumn{3}{c}{\ion{O}{i}} & \multicolumn{3}{c}{\ion{Na}{i}} &  \multicolumn{3}{c}{\ion{Na}{i}} \\
 pm & \multicolumn{3}{c}{777.2} &  \multicolumn{3}{c}{777.4} &  \multicolumn{3}{c}{777.5} & \multicolumn{3}{c}{818.3} &  \multicolumn{3}{c}{819.5} \\
     &  0.5  &  1.0  & 1.5 &  0.5  &  1.0  & 1.5  &  0.5  &  1.0  & 1.5 &  0.5  &  1.0  & 1.5 &  0.5  &  1.0  & 1.5 \\
\hline\noalign{\smallskip}
 0.4  & $-0.001$ & 0.001 & 0.003  & $-0.001$ & 0.001 & 0.003 & $-0.001$ & 0.001 & 0.003  & $-0.022$ & $-0.021$ & $-0.020$ & $-0.022$ &  $-0.021$ & $-0.020$ \\
 0.6  &   0.005  & 0.007 & 0.010  &   0.005  & 0.007 & 0.010 &   0.005  & 0.007 & 0.010  & $-0.023$ & $-0.021$ & $-0.019$ & $-0.022$ &  $-0.021$ & $-0.019$ \\
 0.8  &   0.010  & 0.013 & 0.017  &   0.010  & 0.013 & 0.017 &   0.010  & 0.013 & 0.017  & $-0.023$ & $-0.021$ & $-0.019$ & $-0.023$ &  $-0.021$ & $-0.019$ \\
1.0   &   0.014  & 0.018 & 0.023  &   0.014  & 0.018 & 0.023 &   0.014  & 0.018 & 0.023  & $-0.024$ & $-0.022$ & $-0.018$ & $-0.024$ &  $-0.022$ & $-0.019$ \\
1.2   &   0.018  & 0.022 & 0.029  &   0.018  & 0.022 & 0.028 &   0.018  & 0.023 & 0.029  & $-0.025$ & $-0.022$ & $-0.018$ & $-0.025$ &  $-0.022$ & $-0.018$ \\
1.5   &   0.023  & 0.028 & 0.036  &   0.023  & 0.028 & 0.036 &   0.024  & 0.029 & 0.037  & $-0.027$ & $-0.023$ & $-0.018$ & $-0.026$ &  $-0.022$ & $-0.017$ \\
2.0   &   0.031  & 0.037 & 0.047  &   0.031  & 0.038 & 0.048 &   0.031  & 0.037 & 0.047  & $-0.030$ & $-0.024$ & $-0.017$ & $-0.029$ &  $-0.023$ & $-0.017$ \\
2.5   &   0.037  & 0.045 & 0.057  &   0.037  & 0.045 & 0.056 &   0.037  & 0.045 & 0.056  & $-0.031$ & $-0.024$ & $-0.015$ & $-0.032$ &  $-0.025$ & $-0.016$ \\
3.0   &   0.042  & 0.051 & 0.064  &   0.042  & 0.051 & 0.064 &   0.043  & 0.052 & 0.065  & $-0.034$ & $-0.025$ & $-0.014$ & $-0.036$ &  $-0.027$ & $-0.015$ \\
3.5   &   0.047  & 0.057 & 0.071  &   0.047  & 0.057 & 0.071 &   0.047  & 0.057 & 0.071  & $-0.038$ & $-0.027$ & $-0.013$ & $-0.038$ &  $-0.027$ & $-0.013$ \\
4.0   &   0.051  & 0.061 & 0.077  &   0.051  & 0.062 & 0.078 &   0.051  & 0.062 & 0.077  & $-0.042$ & $-0.029$ & $-0.012$ & $-0.040$ &  $-0.027$ & $-0.011$ \\
4.5   &   0.054  & 0.066 & 0.083  &   0.055  & 0.066 & 0.083 &   0.054  & 0.066 & 0.082  & $-0.046$ & $-0.030$ & $-0.011$ & $-0.044$ &  $-0.029$ & $-0.010$ \\
5.0   &   0.058  & 0.070 & 0.088  &   0.058  & 0.070 & 0.088 &   0.058  & 0.070 & 0.087  & $-0.048$ & $-0.031$ & $-0.009$ & $-0.047$ &  $-0.030$ & $-0.008$ \\
5.5   &   0.061  & 0.074 & 0.093  &   0.061  & 0.074 & 0.093 &   0.061  & 0.073 & 0.092  & $-0.051$ & $-0.031$ & $-0.007$ & $-0.051$ &  $-0.031$ & $-0.007$ \\
6.0   &   0.065  & 0.078 & 0.097  &   0.064  & 0.077 & 0.096 &   0.064  & 0.077 & 0.096  & $-0.053$ & $-0.032$ & $-0.004$ & $-0.054$ &  $-0.033$ & $-0.005$ \\
6.5   &   0.067  & 0.081 & 0.101  &   0.067  & 0.080 & 0.100 &   0.067  & 0.080 & 0.100  & $-0.056$ & $-0.033$ & $-0.002$ & $-0.057$ &  $-0.034$ & $-0.003$ \\
7.0   &   0.070  & 0.083 & 0.104  &   0.069  & 0.083 & 0.103 &   0.070  & 0.083 & 0.104  & $-0.059$ & $-0.033$ &   0.000  & $-0.059$ &  $-0.034$ & $-0.001$ \\
7.5   &   0.072  & 0.086 & 0.106  &   0.071  & 0.085 & 0.106 &   0.072  & 0.086 & 0.107  & $-0.061$ & $-0.034$ &   0.002  & $-0.061$ &  $-0.034$ &   0.002  \\
8.0   &   0.074  & 0.088 & 0.109  &   0.074  & 0.088 & 0.108 &   0.074  & 0.088 & 0.109  & $-0.063$ & $-0.034$ &   0.004  & $-0.062$ &  $-0.034$ &   0.004  \\
8.5   &   0.076  & 0.089 & 0.111  &   0.076  & 0.090 & 0.111 &   0.076  & 0.090 & 0.111  & $-0.064$ & $-0.034$ &   0.006  & $-0.064$ &  $-0.034$ &   0.006  \\
9.0   &   0.077  & 0.091 & 0.112  &   0.078  & 0.092 & 0.114 &   0.078  & 0.092 & 0.113  & $-0.066$ & $-0.034$ &   0.008  & $-0.064$ &  $-0.033$ &   0.009  \\
9.5   &   0.079  & 0.093 & 0.114  &   0.080  & 0.094 & 0.116 &   0.079  & 0.093 & 0.115  & $-0.066$ & $-0.034$ &   0.011  & $-0.064$ &  $-0.033$ &   0.011  \\
10.0  &   0.081  & 0.095 & 0.116  &   0.082  & 0.096 & 0.118 &   0.081  & 0.095 & 0.116  & $-0.066$ & $-0.033$ &   0.013  & $-0.064$ &  $-0.032$ &   0.013  \\
11.0  &   0.084  & 0.098 & 0.119  &   0.084  & 0.098 & 0.120 &   0.083  & 0.097 & 0.118  & $-0.064$ & $-0.031$ &   0.017  & $-0.062$ &  $-0.030$ &   0.017  \\
12.0  &   0.086  & 0.100 & 0.121  &   0.086  & 0.100 & 0.121 &   0.086  & 0.100 & 0.121  & $-0.060$ & $-0.028$ &   0.020  & $-0.061$ &  $-0.028$ &   0.020  \\
13.0  &   0.088  & 0.102 & 0.122  &   0.087  & 0.101 & 0.121 &   0.088  & 0.101 & 0.122  & $-0.056$ & $-0.024$ &   0.023  & $-0.057$ &  $-0.025$ &   0.024  \\
14.0  &   0.089  & 0.102 & 0.123  &   0.089  & 0.102 & 0.122 &   0.089  & 0.103 & 0.123  & $-0.051$ & $-0.021$ &   0.025  & $-0.053$ &  $-0.021$ &   0.026  \\
15.0  &   0.090  & 0.103 & 0.123  &   0.090  & 0.103 & 0.123 &   0.090  & 0.103 & 0.123  & $-0.047$ & $-0.018$ &   0.028  & $-0.048$ &  $-0.018$ &   0.028  \\
16.0  &   \ldots & \ldots & \ldots &   \ldots & \ldots & \ldots &   \ldots & \ldots & \ldots & $-0.043$ & $-0.015$ &   0.029  & $-0.042$ &  $-0.014$ &   0.030  \\
17.0  &   \ldots & \ldots & \ldots &   \ldots & \ldots & \ldots &   \ldots & \ldots & \ldots & $-0.038$ & $-0.011$ &   0.031  & $-0.037$ &  $-0.011$ &   0.030  \\
18.0  &   \ldots & \ldots & \ldots &   \ldots & \ldots & \ldots &   \ldots & \ldots & \ldots & $-0.033$ & $-0.008$ &   0.032  & $-0.033$ &  $-0.008$ &   0.031  \\
19.0  &   \ldots & \ldots & \ldots &   \ldots & \ldots & \ldots &   \ldots & \ldots & \ldots & $-0.029$ & $-0.005$ &   0.033  & $-0.029$ &  $-0.005$ &   0.032  \\
20.0  &   \ldots & \ldots & \ldots &   \ldots & \ldots & \ldots &   \ldots & \ldots & \ldots & $-0.025$ & $-0.003$ &   0.033  & $-0.025$ &  $-0.003$ &   0.033  \\
21.0  &   \ldots & \ldots & \ldots &   \ldots & \ldots & \ldots &   \ldots & \ldots & \ldots & $-0.021$ &   0.000  &   0.033  & $-0.021$ &  $-0.001$ &   0.033  \\
22.0  &   \ldots & \ldots & \ldots &   \ldots & \ldots & \ldots &   \ldots & \ldots & \ldots & $-0.018$ &   0.002  &   0.033  & $-0.018$ &    0.002  &   0.034  \\
23.0  &   \ldots & \ldots & \ldots &   \ldots & \ldots & \ldots &   \ldots & \ldots & \ldots & $-0.015$ &   0.004  &   0.033  & $-0.015$ &    0.004  &   0.034  \\
24.0  &   \ldots & \ldots & \ldots &   \ldots & \ldots & \ldots &   \ldots & \ldots & \ldots & $-0.012$ &   0.005  &   0.033  & $-0.012$ &    0.006  &   0.034  \\
25.0  &   \ldots & \ldots & \ldots &   \ldots & \ldots & \ldots &   \ldots & \ldots & \ldots & $-0.010$ &   0.007  &   0.033  & $-0.009$ &    0.007  &   0.034  \\
26.0  &   \ldots & \ldots & \ldots &   \ldots & \ldots & \ldots &   \ldots & \ldots & \ldots & $-0.008$ &   0.008  &   0.034  & $-0.007$ &    0.009  &   0.034  \\
27.0  &   \ldots & \ldots & \ldots &   \ldots & \ldots & \ldots &   \ldots & \ldots & \ldots & $-0.006$ &   0.009  &   0.034  & $-0.005$ &    0.010  &   0.034  \\
28.0  &   \ldots & \ldots & \ldots &   \ldots & \ldots & \ldots &   \ldots & \ldots & \ldots & $-0.004$ &   0.010  &   0.034  & $-0.003$ &    0.011  &   0.034  \\
29.0  &   \ldots & \ldots & \ldots &   \ldots & \ldots & \ldots &   \ldots & \ldots & \ldots & $-0.002$ &   0.011  &   0.034  & $-0.001$ &    0.012  &   0.034  \\
30.0  &   \ldots & \ldots & \ldots &   \ldots & \ldots & \ldots &   \ldots & \ldots & \ldots & $-0.001$ &   0.013  &   0.034  &   0.000  &    0.013  &   0.034  \\
\hline
\end{tabular}
\end{center}
\end{sidewaystable*}

\begin{sidewaystable*}
\begin{center}
\caption{The same as in Table~\ref{tab:dabut55g40mm00} but for the model atmosphere with $T_{\rm eff} = 5927$~K, $\log g = 4.0$,  $\MoH = 0.0$.
\label{tab:dabut59g40mm00}}
  \begin{tabular}{rccccccccccccccc}
  \hline\hline
\noalign{\smallskip}
{\it EW}, & \multicolumn{3}{c}{\ion{O}{i}} & \multicolumn{3}{c}{\ion{O}{i}}  & \multicolumn{3}{c}{\ion{O}{i}} & \multicolumn{3}{c}{\ion{Na}{i}} &  \multicolumn{3}{c}{\ion{Na}{i}} \\
 pm & \multicolumn{3}{c}{777.2} &  \multicolumn{3}{c}{777.4} &  \multicolumn{3}{c}{777.5} & \multicolumn{3}{c}{818.3} &  \multicolumn{3}{c}{819.5} \\
     &  0.5  &  1.0  & 1.5 &  0.5  &  1.0  & 1.5  &  0.5  &  1.0  & 1.5 &  0.5  &  1.0  & 1.5 &  0.5  &  1.0  & 1.5 \\
\hline\noalign{\smallskip}
 0.4  & $-0.013$ & $-0.011$ & $-0.008$ & $-0.012$  & $-0.011$ & $-0.008$ & $-0.012$  & $-0.011$ & $-0.008$  &   0.021  & 0.022 & 0.024 &   0.021  &  0.022 & 0.023 \\
 0.6  & $-0.010$ & $-0.008$ & $-0.004$ & $-0.010$  & $-0.007$ & $-0.004$ & $-0.011$  & $-0.008$ & $-0.004$  &   0.020  & 0.022 & 0.024 &   0.020  &  0.022 & 0.024 \\
 0.8  & $-0.008$ & $-0.005$ &   0.000  & $-0.009$  & $-0.005$ &   0.000  & $-0.008$  & $-0.005$ &   0.000   &   0.020  & 0.022 & 0.025 &   0.019  &  0.022 & 0.025 \\
 1.0  & $-0.007$ & $-0.002$ &   0.004  & $-0.007$  & $-0.002$ &   0.004  & $-0.007$  & $-0.002$ &   0.004   &   0.019  & 0.022 & 0.026 &   0.018  &  0.021 & 0.025 \\
 1.2  & $-0.005$ &   0.000  &   0.008  & $-0.005$  &   0.000  &   0.008  & $-0.005$  &   0.000  &   0.008   &   0.018  & 0.022 & 0.027 &   0.018  &  0.021 & 0.026 \\
 1.5  & $-0.003$ &   0.004  &   0.013  & $-0.003$  &   0.004  &   0.013  & $-0.003$  &   0.004  &   0.014   &   0.016  & 0.021 & 0.027 &   0.017  &  0.022 & 0.027 \\
 2.0  &   0.000  &   0.009  &   0.021  &   0.000   &   0.009  &   0.022  &   0.000   &   0.009  &   0.021   &   0.014  & 0.021 & 0.029 &   0.015  &  0.022 & 0.030 \\
 2.5  &   0.003  &   0.014  &   0.029  &   0.003   &   0.014  &   0.029  &   0.003   &   0.013  &   0.029   &   0.012  & 0.021 & 0.032 &   0.011  &  0.020 & 0.031 \\
 3.0  &   0.005  &   0.018  &   0.036  &   0.005   &   0.018  &   0.036  &   0.005   &   0.018  &   0.036   &   0.010  & 0.021 & 0.036 &   0.008  &  0.019 & 0.033 \\
 3.5  &   0.007  &   0.022  &   0.042  &   0.007   &   0.022  &   0.042  &   0.007   &   0.022  &   0.043   &   0.006  & 0.020 & 0.037 &   0.006  &  0.019 & 0.036 \\
 4.0  &   0.009  &   0.025  &   0.048  &   0.009   &   0.025  &   0.049  &   0.009   &   0.025  &   0.048   &   0.002  & 0.018 & 0.038 &   0.003  &  0.020 & 0.040 \\
 4.5  &   0.011  &   0.029  &   0.054  &   0.012   &   0.029  &   0.054  &   0.011   &   0.029  &   0.054   & $-0.002$ & 0.017 & 0.041 &   0.000  &  0.020 & 0.045 \\
 5.0  &   0.014  &   0.032  &   0.060  &   0.014   &   0.032  &   0.059  &   0.013   &   0.032  &   0.059   & $-0.005$ & 0.017 & 0.045 & $-0.004$ &  0.019 & 0.048 \\
 5.5  &   0.016  &   0.036  &   0.065  &   0.016   &   0.035  &   0.064  &   0.016   &   0.035  &   0.064   & $-0.009$ & 0.018 & 0.050 & $-0.008$ &  0.018 & 0.050 \\
 6.0  &   0.018  &   0.039  &   0.069  &   0.018   &   0.038  &   0.068  &   0.018   &   0.039  &   0.069   & $-0.012$ & 0.018 & 0.055 & $-0.013$ &  0.017 & 0.053 \\
 6.5  &   0.020  &   0.042  &   0.073  &   0.020   &   0.041  &   0.072  &   0.020   &   0.042  &   0.073   & $-0.015$ & 0.019 & 0.061 & $-0.017$ &  0.016 & 0.056 \\
 7.0  &   0.022  &   0.044  &   0.077  &   0.022   &   0.044  &   0.076  &   0.022   &   0.045  &   0.077   & $-0.018$ & 0.019 & 0.065 & $-0.020$ &  0.016 & 0.061 \\
 7.5  &   0.024  &   0.046  &   0.080  &   0.024   &   0.047  &   0.080  &   0.025   &   0.047  &   0.080   & $-0.022$ & 0.018 & 0.068 & $-0.023$ &  0.016 & 0.066 \\
 8.0  &   0.026  &   0.048  &   0.082  &   0.027   &   0.049  &   0.083  &   0.026   &   0.049  &   0.083   & $-0.025$ & 0.018 & 0.072 & $-0.025$ &  0.017 & 0.071 \\
 8.5  &   0.028  &   0.051  &   0.085  &   0.029   &   0.052  &   0.086  &   0.028   &   0.051  &   0.086   & $-0.027$ & 0.018 & 0.076 & $-0.027$ &  0.019 & 0.077 \\
 9.0  &   0.030  &   0.053  &   0.088  &   0.031   &   0.054  &   0.089  &   0.030   &   0.053  &   0.087   & $-0.029$ & 0.018 & 0.080 & $-0.028$ &  0.020 & 0.083 \\
 9.5  &   0.032  &   0.056  &   0.091  &   0.033   &   0.056  &   0.091  &   0.032   &   0.055  &   0.089   & $-0.030$ & 0.019 & 0.084 & $-0.029$ &  0.022 & 0.088 \\
10.0  &   0.034  &   0.058  &   0.093  &   0.034   &   0.058  &   0.093  &   0.034   &   0.057  &   0.092   & $-0.029$ & 0.020 & 0.088 & $-0.029$ &  0.023 & 0.092 \\
11.0  &   0.038  &   0.061  &   0.096  &   0.037   &   0.060  &   0.095  &   0.038   &   0.061  &   0.095   & $-0.028$ & 0.025 & 0.098 & $-0.027$ &  0.025 & 0.099 \\
12.0  &   0.041  &   0.063  &   0.097  &   0.041   &   0.063  &   0.097  &   0.041   &   0.064  &   0.098   & $-0.023$ & 0.030 & 0.106 & $-0.023$ &  0.028 & 0.103 \\
13.0  &   0.043  &   0.065  &   0.098  &   0.044   &   0.066  &   0.100  &   0.044   &   0.066  &   0.099   & $-0.017$ & 0.035 & 0.112 & $-0.018$ &  0.033 & 0.107 \\
14.0  &   0.045  &   0.066  &   0.098  &   0.047   &   0.068  &   0.101  &   0.046   &   0.067  &   0.100   & $-0.011$ & 0.039 & 0.115 & $-0.012$ &  0.039 & 0.113 \\
15.0  &   0.048  &   0.068  &   0.099  &   0.049   &   0.070  &   0.102  &   0.048   &   0.068  &   0.100   & $-0.005$ & 0.042 & 0.115 & $-0.004$ &  0.044 & 0.118 \\
16.0  &   \ldots &   \ldots &   \ldots &    \ldots &   \ldots &   \ldots &    \ldots &   \ldots &   \ldots  &   0.002  & 0.047 & 0.116 &   0.003  &  0.049 & 0.120 \\
17.0  &   \ldots &   \ldots &   \ldots &    \ldots &   \ldots &   \ldots &    \ldots &   \ldots &   \ldots  &   0.009  & 0.052 & 0.119 &   0.010  &  0.052 & 0.120 \\
18.0  &   \ldots &   \ldots &   \ldots &    \ldots &   \ldots &   \ldots &    \ldots &   \ldots &   \ldots  &   0.015  & 0.056 & 0.120 &   0.016  &  0.055 & 0.118 \\
19.0  &   \ldots &   \ldots &   \ldots &    \ldots &   \ldots &   \ldots &    \ldots &   \ldots &   \ldots  &   0.022  & 0.060 & 0.121 &   0.021  &  0.057 & 0.116 \\
20.0  &   \ldots &   \ldots &   \ldots &    \ldots &   \ldots &   \ldots &    \ldots &   \ldots &   \ldots  &   0.027  & 0.063 & 0.120 &   0.026  &  0.060 & 0.116 \\
21.0  &   \ldots &   \ldots &   \ldots &    \ldots &   \ldots &   \ldots &    \ldots &   \ldots &   \ldots  &   0.032  & 0.065 & 0.119 &   0.030  &  0.063 & 0.115 \\
22.0  &   \ldots &   \ldots &   \ldots &    \ldots &   \ldots &   \ldots &    \ldots &   \ldots &   \ldots  &   0.035  & 0.066 & 0.116 &   0.034  &  0.065 & 0.115 \\
23.0  &   \ldots &   \ldots &   \ldots &    \ldots &   \ldots &   \ldots &    \ldots &   \ldots &   \ldots  &   0.038  & 0.067 & 0.114 &   0.038  &  0.067 & 0.114 \\
24.0  &   \ldots &   \ldots &   \ldots &    \ldots &   \ldots &   \ldots &    \ldots &   \ldots &   \ldots  &   0.041  & 0.067 & 0.111 &   0.042  &  0.069 & 0.113 \\
25.0  &   \ldots &   \ldots &   \ldots &    \ldots &   \ldots &   \ldots &    \ldots &   \ldots &   \ldots  &   0.043  & 0.068 & 0.109 &   0.046  &  0.071 & 0.113 \\
26.0  &   \ldots &   \ldots &   \ldots &    \ldots &   \ldots &   \ldots &    \ldots &   \ldots &   \ldots  &   0.045  & 0.069 & 0.108 &   0.049  &  0.073 & 0.112 \\
27.0  &   \ldots &   \ldots &   \ldots &    \ldots &   \ldots &   \ldots &    \ldots &   \ldots &   \ldots  &   0.048  & 0.070 & 0.106 &   0.051  &  0.074 & 0.111 \\
28.0  &   \ldots &   \ldots &   \ldots &    \ldots &   \ldots &   \ldots &    \ldots &   \ldots &   \ldots  &   0.050  & 0.071 & 0.106 &   0.053  &  0.075 & 0.110 \\
29.0  &   \ldots &   \ldots &   \ldots &    \ldots &   \ldots &   \ldots &    \ldots &   \ldots &   \ldots  &   0.052  & 0.072 & 0.105 &   0.055  &  0.076 & 0.109 \\
30.0  &   \ldots &   \ldots &   \ldots &    \ldots &   \ldots &   \ldots &    \ldots &   \ldots &   \ldots  &   0.054  & 0.074 & 0.105 &   0.056  &  0.076 & 0.108 \\
\hline
\end{tabular}
\end{center}
\end{sidewaystable*}

\begin{sidewaystable*}
\begin{center}
\caption{The same as in Table~\ref{tab:dabut55g40mm00} but for the model atmosphere with $T_{\rm eff} = 5850$~K, $\log g = 4.0$,  $\MoH = -1.0$.
\label{tab:dabut59g40mm10}}
  \begin{tabular}{rccccccccccccccc}
  \hline\hline
\noalign{\smallskip}
{\it EW}, & \multicolumn{3}{c}{\ion{O}{i}} & \multicolumn{3}{c}{\ion{O}{i}}  & \multicolumn{3}{c}{\ion{O}{i}} & \multicolumn{3}{c}{\ion{Na}{i}} &  \multicolumn{3}{c}{\ion{Na}{i}} \\
 pm & \multicolumn{3}{c}{777.2} &  \multicolumn{3}{c}{777.4} &  \multicolumn{3}{c}{777.5} & \multicolumn{3}{c}{818.3} &  \multicolumn{3}{c}{819.5} \\
     &  0.5  &  1.0  & 1.5 &  0.5  &  1.0  & 1.5  &  0.5  &  1.0  & 1.5 &  0.5  &  1.0  & 1.5 &  0.5  &  1.0  & 1.5 \\
\hline\noalign{\smallskip}
 0.4  & $-0.021$ & $-0.020$ & $-0.018$ & $-0.022$  & $-0.020$ & $-0.018$ & $-0.021$  & $-0.020$ & $-0.018$  & $-0.003$ & $-0.002$ & $-0.001$ & $-0.003$ &  $-0.002$ & $-0.001$ \\
 0.6  & $-0.018$ & $-0.016$ & $-0.013$ & $-0.018$  & $-0.015$ & $-0.012$ & $-0.017$  & $-0.015$ & $-0.012$  & $-0.004$ & $-0.003$ & $-0.001$ & $-0.004$ &  $-0.002$ & $-0.001$ \\
 0.8  & $-0.014$ & $-0.011$ & $-0.007$ & $-0.014$  & $-0.011$ & $-0.007$ & $-0.014$  & $-0.011$ & $-0.007$  & $-0.005$ & $-0.003$ &   0.000  & $-0.005$ &  $-0.003$ &   0.000  \\
 1.0  & $-0.010$ & $-0.007$ & $-0.002$ & $-0.010$  & $-0.007$ & $-0.002$ & $-0.010$  & $-0.007$ & $-0.002$  & $-0.006$ & $-0.003$ &   0.000  & $-0.006$ &  $-0.003$ &   0.000  \\
 1.2  & $-0.007$ & $-0.003$ &   0.003  & $-0.007$  &   0.003  &   0.003  & $-0.007$  & $-0.003$ &   0.003   & $-0.007$ & $-0.004$ &   0.000  & $-0.007$ &  $-0.004$ &   0.000  \\
 1.5  & $-0.002$ &   0.003  &   0.011  & $-0.002$  &   0.003  &   0.010  & $-0.002$  &   0.003  &   0.010   & $-0.009$ & $-0.005$ &   0.001  & $-0.008$ &  $-0.004$ &   0.001  \\
 2.0  &   0.005  &   0.012  &   0.021  &   0.005   &   0.012  &   0.021  &   0.005   &   0.012  &   0.022   & $-0.011$ & $-0.006$ &   0.002  & $-0.011$ &  $-0.006$ &   0.002  \\
 2.5  &   0.012  &   0.020  &   0.031  &   0.012   &   0.020  &   0.031  &   0.012   &   0.020  &   0.032   & $-0.014$ & $-0.007$ &   0.003  & $-0.015$ &  $-0.007$ &   0.003  \\
 3.0  &   0.018  &   0.027  &   0.041  &   0.018   &   0.028  &   0.041  &   0.018   &   0.027  &   0.041   & $-0.018$ & $-0.008$ &   0.004  & $-0.018$ &  $-0.008$ &   0.004  \\
 3.5  &   0.024  &   0.034  &   0.050  &   0.024   &   0.035  &   0.050  &   0.024   &   0.034  &   0.049   & $-0.022$ & $-0.010$ &   0.005  & $-0.022$ &  $-0.010$ &   0.004  \\
 4.0  &   0.029  &   0.041  &   0.058  &   0.029   &   0.041  &   0.058  &   0.029   &   0.041  &   0.058   & $-0.025$ & $-0.011$ &   0.006  & $-0.026$ &  $-0.012$ &   0.006  \\
 4.5  &   0.034  &   0.047  &   0.065  &   0.034   &   0.047  &   0.065  &   0.034   &   0.047  &   0.066   & $-0.029$ & $-0.013$ &   0.007  & $-0.030$ &  $-0.013$ &   0.007  \\
 5.0  &   0.039  &   0.052  &   0.072  &   0.039   &   0.053  &   0.073  &   0.039   &   0.053  &   0.073   & $-0.034$ & $-0.015$ &   0.008  & $-0.034$ &  $-0.015$ &   0.009  \\
 5.5  &   0.043  &   0.058  &   0.079  &   0.043   &   0.058  &   0.079  &   0.044   &   0.058  &   0.079   & $-0.038$ & $-0.017$ &   0.009  & $-0.037$ &  $-0.017$ &   0.010  \\
 6.0  &   0.047  &   0.063  &   0.085  &   0.048   &   0.063  &   0.085  &   0.048   &   0.063  &   0.085   & $-0.042$ & $-0.019$ &   0.011  & $-0.041$ &  $-0.018$ &   0.012  \\
 6.5  &   0.051  &   0.067  &   0.090  &   0.052   &   0.068  &   0.091  &   0.052   &   0.068  &   0.091   & $-0.046$ & $-0.020$ &   0.013  & $-0.045$ &  $-0.020$ &   0.013  \\
 7.0  &   0.055  &   0.071  &   0.095  &   0.056   &   0.073  &   0.097  &   0.055   &   0.072  &   0.096   & $-0.049$ & $-0.021$ &   0.015  & $-0.049$ &  $-0.021$ &   0.014  \\
 7.5  &   0.059  &   0.076  &   0.100  &   0.060   &   0.077  &   0.101  &   0.059   &   0.075  &   0.100   & $-0.052$ & $-0.022$ &   0.017  & $-0.052$ &  $-0.023$ &   0.016  \\
 8.0  &   0.063  &   0.080  &   0.105  &   0.063   &   0.080  &   0.106  &   0.062   &   0.079  &   0.104   & $-0.054$ & $-0.023$ &   0.019  & $-0.054$ &  $-0.024$ &   0.018  \\
 8.5  &   0.066  &   0.083  &   0.109  &   0.066   &   0.083  &   0.109  &   0.065   &   0.082  &   0.108   & $-0.056$ & $-0.023$ &   0.021  & $-0.056$ &  $-0.024$ &   0.020  \\
 9.0  &   0.069  &   0.086  &   0.112  &   0.068   &   0.085  &   0.112  &   0.068   &   0.085  &   0.112   & $-0.057$ & $-0.023$ &   0.023  & $-0.057$ &  $-0.024$ &   0.022  \\
 9.5  &   0.071  &   0.089  &   0.115  &   0.070   &   0.087  &   0.114  &   0.071   &   0.088  &   0.115   & $-0.058$ & $-0.023$ &   0.025  & $-0.058$ &  $-0.023$ &   0.024  \\
10.0  &   0.073  &   0.091  &   0.117  &   0.072   &   0.089  &   0.116  &   0.073   &   0.091  &   0.117   & $-0.058$ & $-0.023$ &   0.027  & $-0.058$ &  $-0.023$ &   0.027  \\
11.0  &   0.076  &   0.094  &   0.120  &   0.076   &   0.093  &   0.120  &   0.077   &   0.094  &   0.121   & $-0.056$ & $-0.021$ &   0.030  & $-0.056$ &  $-0.020$ &   0.032  \\
12.0  &   0.078  &   0.096  &   0.122  &   0.079   &   0.096  &   0.123  &   0.079   &   0.096  &   0.123   & $-0.052$ & $-0.017$ &   0.034  & $-0.052$ &  $-0.016$ &   0.036  \\
13.0  &   0.081  &   0.098  &   0.124  &   0.081   &   0.098  &   0.125  &   0.081   &   0.098  &   0.124   & $-0.047$ & $-0.013$ &   0.039  & $-0.046$ &  $-0.012$ &   0.040  \\
14.0  &   0.083  &   0.100  &   0.125  &   0.083   &   0.100  &   0.125  &   0.082   &   0.099  &   0.125   & $-0.040$ & $-0.007$ &   0.043  & $-0.039$ &  $-0.007$ &   0.043  \\
15.0  &   0.084  &   0.101  &   0.126  &   0.084   &   0.100  &   0.125  &   0.084   &   0.100  &   0.125   & $-0.033$ & $-0.002$ &   0.047  & $-0.033$ &  $-0.003$ &   0.045  \\
16.0  &  \ldots  &   \ldots &   \ldots &    \ldots &   \ldots &   \ldots &    \ldots &   \ldots &   \ldots  & $-0.026$ &   0.004  &   0.050  & $-0.027$ &    0.002  &   0.048  \\
17.0  &  \ldots  &   \ldots &   \ldots &    \ldots &   \ldots &   \ldots &    \ldots &   \ldots &   \ldots  & $-0.019$ &   0.008  &   0.052  & $-0.020$ &    0.007  &   0.051  \\
18.0  &  \ldots  &   \ldots &   \ldots &    \ldots &   \ldots &   \ldots &    \ldots &   \ldots &   \ldots  & $-0.013$ &   0.012  &   0.053  & $-0.014$ &    0.012  &   0.054  \\
19.0  &  \ldots  &   \ldots &   \ldots &    \ldots &   \ldots &   \ldots &    \ldots &   \ldots &   \ldots  & $-0.008$ &   0.016  &   0.055  & $-0.007$ &    0.017  &   0.056  \\
20.0  &  \ldots  &   \ldots &   \ldots &    \ldots &   \ldots &   \ldots &    \ldots &   \ldots &   \ldots  & $-0.003$ &   0.020  &   0.057  & $-0.002$ &    0.021  &   0.058  \\
21.0  &  \ldots  &   \ldots &   \ldots &    \ldots &   \ldots &   \ldots &    \ldots &   \ldots &   \ldots  &   0.002  &   0.024  &   0.058  &   0.004  &    0.025  &   0.060  \\
22.0  &  \ldots  &   \ldots &   \ldots &    \ldots &   \ldots &   \ldots &    \ldots &   \ldots &   \ldots  &   0.007  &   0.027  &   0.060  &   0.008  &    0.028  &   0.061  \\
23.0  &  \ldots  &   \ldots &   \ldots &    \ldots &   \ldots &   \ldots &    \ldots &   \ldots &   \ldots  &   0.011  &   0.030  &   0.061  &   0.012  &    0.031  &   0.061  \\
24.0  &  \ldots  &   \ldots &   \ldots &    \ldots &   \ldots &   \ldots &    \ldots &   \ldots &   \ldots  &   0.015  &   0.033  &   0.062  &   0.015  &    0.033  &   0.062  \\
25.0  &  \ldots  &   \ldots &   \ldots &    \ldots &   \ldots &   \ldots &    \ldots &   \ldots &   \ldots  &   0.019  &   0.036  &   0.063  &   0.018  &    0.035  &   0.062  \\
26.0  &  \ldots  &   \ldots &   \ldots &    \ldots &   \ldots &   \ldots &    \ldots &   \ldots &   \ldots  &   0.022  &   0.038  &   0.064  &   0.021  &    0.037  &   0.063  \\
27.0  &  \ldots  &   \ldots &   \ldots &    \ldots &   \ldots &   \ldots &    \ldots &   \ldots &   \ldots  &   0.025  &   0.040  &   0.065  &   0.024  &    0.039  &   0.063  \\
28.0  &  \ldots  &   \ldots &   \ldots &    \ldots &   \ldots &   \ldots &    \ldots &   \ldots &   \ldots  &   0.028  &   0.042  &   0.066  &   0.026  &    0.040  &   0.064  \\
29.0  &  \ldots  &   \ldots &   \ldots &    \ldots &   \ldots &   \ldots &    \ldots &   \ldots &   \ldots  &   0.030  &   0.044  &   0.066  &   0.028  &    0.042  &   0.065  \\
30.0  &  \ldots  &   \ldots &   \ldots &    \ldots &   \ldots &   \ldots &    \ldots &   \ldots &   \ldots  &   0.032  &   0.045  &   0.066  &   0.030  &    0.044  &   0.065  \\
\hline
\end{tabular}
\end{center}
\end{sidewaystable*}

\end{appendix}

\end{document}